\documentclass{emulateapj}

\usepackage{amssymb}
\shorttitle{Optically ``Invisible'' Sources in Bo\"{o}tes}
\shortauthors{Higdon et al.}

\begin{document}
\slugcomment{to appear in the Astrophysical Journal}

\title{Radio and Infrared Selected Optically Invisible Sources In The Bo\"otes NDWFS}

\author{J.~L.~Higdon\altaffilmark{1}, 
	S.~J.~U. Higdon\altaffilmark{1},
	S.~P.~Willner\altaffilmark{2},
 	M.~J.~I. Brown\altaffilmark{3},
	D.~Stern\altaffilmark{4},
	E.~Le~Floc'h\altaffilmark{5,6}, and
	P.~Eisenhardt\altaffilmark{4}
   }

\altaffiltext{1}{Department of Physics, Georgia Southern University, Statesboro, GA 30460}
\altaffiltext{2}{Harvard-Smithsonian Center for Astrophysics, 60 Garden
                 Street, Cambridge, MA 02138}
\altaffiltext{3}{School of Physics, Monash University, Clayton, Victoria 3800, Australia }
\altaffiltext{4}{Jet Propulsion Laboratory, Caltech, MC 169-327, 4800
                 Oak Grove Drive, Pasadena, CA 91109}
\altaffiltext{5}{Steward Observatory, University of Arizona, Tucson, AZ 85721}
\altaffiltext{6}{Institute for Astronomy, University of Hawaii, 
                   2680 Woodlawn Drive, Honolulu HI 96822}

\begin{abstract}
We have combined data from the NOAO Deep Wide-Field Survey in
Bo\"{o}tes and the {\em Spitzer Space Telescope} to determine 
basic properties  for sixteen optically ``invisible'' MIPS 24 $\mu$m (OIMS) and 
thirty-five optically ``invisible'' radio 
(OIRS) sources, including their spectral energy distributions (SED) and luminosities.  
Most OIMSs possess steep power-law SEDs over $\lambda_{\rm rest} = 
1-10 ~\mu m$, indicating the presence of obscured AGN in agreement with
{\em Spitzer} spectroscopy. These objects are extremely luminous at
rest-frame near and mid-IR ($\nu L_{\nu}(5 ~\mu m) \approx 10^{38}-10^{39}$ W),
consistent with accretion near the Eddington limit and further implying that they host
buried QSOs. The majority of the IRAC detected OIRSs have flat 3.6 to 24 $\mu$m SEDs, 
implying comparable emission from stellar photospheres and hot AGN illuminated dust.
This may reflect relatively small amounts of dust close to the central engine or current 
low mass accretion rates. A small subset of OIRSs 
appear to be starburst dominated with photometric redshifts from 1.0 to 4.5. 
The OIMSs and OIRSs with significant starburst components
have similar L$_{\rm K}$ and stellar masses ($M_* \approx 10^{11}$ M$_{\odot}$)
assuming minimal AGN contribution.  Roughly half of the 
OIRSs are not detected by {\it Spitzer's} IRAC or MIPS. These are most likely 
$z \ga 2$ radio galaxies.  The IRAC detected OIRSs are more likely 
than OIMSs to appear non point-like in the 3.6 $\mu m$ and 4.5 $\mu$m images,
suggesting that interactions play a role in triggering their activity. 
The AGN powered OIMSs may represent sub-millimeter galaxies making the transition
from starburst to accretion dominance in their evolution to current epoch massive ellipticals.
\end{abstract}


\section{Introduction}

Sensitive multi-band surveys in the optical and near-infrared portion of
the electromagnetic spectrum have proven extremely successful in
identifying large and diverse galaxy populations in the high-redshift universe.
Deep images obtained through optical {\it U$_n$GR} filters, for example, 
readily find $z \ga 3$ galaxies by the presence of a strong 912 \AA ~Lyman break (e.g.,
Steidel et al. 1996).  Lyman Break Galaxies (LBGs) tend to be massive galaxies
with star formation rates (SFR) between $\approx 4-25$ M$_{\odot}$ yr$^{-1}$ and 
relatively low dust obscuration.  Similarly, unobscured star forming disk (S-BzK galaxies)
and passively evolving spheroidal galaxies at $z \sim 1.4 - 3$ can be discerned from 
the myriad background sources by their optical/near-infrared colors or other
magnitude-color criteria (see, e.g., Daddi et al. 2004; Adelberger et al. 2004).
The S-BzK galaxies in particular tend to possess large stellar masses ($M_* \approx 10^{11}$ 
M$_{\odot}$) with SFR $\approx 200$ M$_{\odot}$ yr$^{-1}$.
Nevertheless, it had been long suspected that a substantial fraction of luminous $z > 1$ galaxies
powered by star formation or accretion onto super-massive black holes would not
be identified using these and related techniques because of heavy dust obscuration, but would turn up 
instead at infrared, sub-millimeter, and radio wavelengths (see, e.g., Chary \& 
Elbaz 2001 and references within). Indeed, deep but small-area surveys at sub-millimeter 
wavelengths by SCUBA (e.g., Smail et al. 1997) revealed a population of extremely dusty
galaxies at $z \approx 1-3$, whose high luminosities are generally though to 
originate primarily in star formation (SFR $\ga 1000$ M$_{\odot}$ yr$^{-1}$), though a 
small fraction may be AGN dominated. Galaxies powered by starbursts or active nuclei (AGN) 
can also be strong emitters at radio wavelengths through either large populations
of  supernova remnants or accretion disk phenomena. Even though this emission is unaffected
by obscuration, roughly $10-15\%$ of compact radio sources identified in deep radio surveys
have either extremely faint optical counterparts or none at all (e.g., Richards et al. 1999; 
Fomalont et al. 2002), and may represent
distant luminous obscured starburst or AGN powered galaxies.  Such objects make a 
still unknown contribution to the luminosity, chemical, 
and accretion history of the universe. Determining an accurate census of heavily obscured
sources - their redshifts, luminosities, space densities, and dominant power source - is
one of the key issues in observational astrophysics and a primary objective of the
{\it Spitzer Space Telescope} 
({\em Spitzer}; Werner et al. 2004).\footnotemark[1] \footnotetext[1]{The 
{\em Spitzer Space Telescope} is operated by JPL, California Institute of 
Technology for the National Aeronautics and Space Administration.  
Information on {\em Spitzer} can be found at http:$//$ssc.spitzer.caltech.edu$/$.}

A number of ambitious large area multi-wavelength surveys incorporating {\it Spitzer} data have
been conducted to date. These include the {\it Great Observatories Origins Deep Survey} (GOODS),
the {\it Spitzer Wide-area InfraRed Extra-galactic} survey (SWIRE), and the {\it First
Look Survey} (FLS). 
This paper is concerned with the nature of
infrared and radio selected sources with extremely faint optical counterparts originally selected 
from the {\it  NOAO Deep Wide-Field Survey} (NDWFS) in Bo\"{o}tes (Jannuzi \& Dey 1999)\footnotemark[2]
with the expectation that they are either heavily extincted, at high redshift, or both.
\footnotetext[2]{The NOAO Deep 
Wide-Field Survey is supported by the National Optical Astronomy Observatory, 
which is operated by AURA, Inc., under a cooperative agreement with the
National Science Foundation. Information on the NDWFS can 
be found at http:$//$www.noao.edu$/$noao$/$noaodeep$/$.} Higdon et al. (2005; hereafter, Higdon05) 
identified thirty-six {\em Optically ``Invisible'' Radio Sources} 
(hereafter, OIRSs) out of 377 compact or unresolved radio sources found in a VLA A-array 
20 cm survey covering 0.5 deg$^{2}$ in the NDWFS Bo\"{o}tes field. These observations
reached a  flux density limit of $\approx 80$ $\mu$Jy ($5 \sigma$) 
at the three overlapping pointing centers. To be considered optically ``invisible'', 
a radio source must have no $B_W$, $R$, or $I$ counterpart within 
a 1.5\arcsec~ radius. This corresponds to limiting $3 \sigma$ magnitudes 
of approximately 26.9, 25.6, and 24.9 (Vega; 2\arcsec ~aperture) respectively, 
with the precise value depending upon location within the optical survey. 
The NDWFS region was surveyed at 24 $\mu$m and 70 $\mu$m with
the Multiband Imaging Photometer for {\em Spitzer} (MIPS, Rieke et al. 2004).
Analysis of the OIRS's {\it q}-parameter led Higdon05 
to conclude that they are a population powered by relatively 
unobscured radio-loud active nuclei.\footnotemark[3] While none of the OIRSs have measured
redshifts, Higdon05 argued that they are likely to be at $z > 1$ based on the 
faint optical limits set by the NDWFS survey. 
\footnotetext[3]{{\it q~}$ \equiv log(F_{24 \mu m}/F_{20 cm})$. Systems
powered by star formation or radio-quiet active nuclei possess
{\it q} $= 0.5-1.1$.  Smaller, and in particular, negative values of {\it q}
indicate the presence of increasingly radio-loud active nuclei
(Appleton et al. 2004).} An independently defined sample of seven optically faint ($R \ge$ 24.5) 
and ten optically ``invisible'' sources from the NDWFS Bo\"{o}tes region,
subject to the additional constraints that $F_{\rm 24 \mu m} >$ 0.75 mJy and
$\nu f_{\nu}(24 \mu m)/ \nu f_{\nu}(0.8 \mu m) \ga$ 100,
was observed using {\it Spitzer's} Infrared Spectrometer (IRS, Houck et al. 2004).
These {\em Optically ``Invisible'' MIPS Sources} (hereafter, OIMSs) are a high-z
population, with $z \sim 1.6-2.7$. Comparisons with mid-IR spectra of local starburst
and AGN dominated galaxies suggested that the primary energy source in thirteen of
the seventeen OIMSs is a heavily obscured active nucleus (Houck et al. 2005, hereafter Houck05).
All seventeen have inferred $8 - 1000 ~\mu m$ luminosities of $\approx$10$^{13}$~ L$_{\odot}$,
placing them in the ``hyper''-luminous class.

It is not known how the radio and infrared selected ``invisible'' 
populations in Higdon05 and Houck05 are related, apart from the 
fact that the majority of both appear to possess powerful AGN. We do not 
know, for example, if the OIRS population lies at systematically higher redshifts 
than the OIMSs with similar L$_{\rm IR}$, or for that matter, if they 
are a sub-L$^{*}$ population at $z \sim 0.5$. Key to addressing these issues
is estimating photometric redshifts for the OIRSs, which are generally too faint 
for {\em Spitzer's} IRS. In this paper, we present
results obtained by combining optical $B_W$, $R$, and $I$, and MIPS
24 $\mu$m, 70 $\mu$m,  and 160 $\mu$m flux densities (or limits) with
data from {\em Spitzer's} Infrared Array Camera (IRAC, Fazio et al. 2004)
Shallow Survey of the NDWFS Bo\"{o}tes field (Eisenhardt et al. 2004). 
The IRAC Shallow Survey covers 8.5 deg$^{2}$ in four bands centered at
3.6, 4.5, 5.8, \& 8.0 $\mu$m. For $z \approx 1-3$, these bands measure emission 
in the rest-frame near-infrared.  Because this wavelength regime 
is much less sensitive to extinction relative to the optical, the
IRAC data can constrain the evolved stellar content of galaxies
at this epoch. Photometric redshifts are also possible by
virtue of the rest-frame 1.6 $\mu$m peak arising from the H$^{-}$ 
opacity minimum in the photospheres of evolved stars.  
It is our aim to derive basic properties for both optically ``invisible'' 
source populations, including rest-frame near-infrared
luminosities and SEDs, and to determine, for example, if their rest-frame
near-infrared emission is dominated by starlight or an AGN. Only then
would it be possible to relate OIMSs and OIRSs to other high-z populations.
It is also our intent to use the IRAC images to explore the near environments
of OIMSs and OIRSs for additional clues to their nature and evolution.

A brief description of the extraction of IRAC flux densities
for the infrared and radio selected ``invisible'' sources  is given 
in $\S$2. Photometric redshift estimates for selected OIRSs are 
presented in $\S$3, along with determinations of the average rest-frame
SEDs,  luminosities, and near environments 
of the OIMS and  OIRS samples detected with IRAC. The nature of
these two populations is discussed in $\S$4. 
These results are summarized in $\S5$. Throughout this paper we will refer
to the OIRSs and OIMSs by number (i.e., OIRS \#97 or OIMS \#13),
corresponding to their entries in Table  2 of Higdon05 and
Table 1 of Houck05. Source coordinates can be found from those papers
directly, or through their SIMBAD designations, [HHW2005]~\# and [HSW2005]~\#
for the OIRSs and OIMSs respectively.
We assume a flat $\Lambda$CDM cosmology, with $\Omega_{\rm M}$=0.27, 
$\Omega_{\rm \Lambda}$=0.73, and a Hubble constant of 71 km s$^{-1}$ Mpc$^{-1}$.

   \section{Observations And Source Extraction}

Flux densities or limits at 3.6, 4.5, 5.8, and 8.0 $\mu$m
for the OIRSs and OIMSs were extracted
from the final IRAC Shallow Survey mosaics (pipeline version S11) using SExtractor
(Bertin \& Arnouts 1996) with 4.8\arcsec~ diameter apertures.
All sources are nearly point-like, and point-source aperture
corrections were applied.  The 3.6 $\mu$m images were used for matching the
radio and 24 $\mu$m selected source
positions since they had the highest point source sensitivity 
($F_{\rm 3.6 \mu m}$ = 6.4 $\mu$Jy ($5 \sigma$) in a 4\arcsec ~diameter 
aperture) and angular resolution ($1.7''$ FWHM). To qualify as a ``match'' the 
coordinates of an OIMS or OIRS had to agree with an IRAC 3.6 $\mu$m
source to within $2\arcsec$. We decided upon this match radius after
comparing the radio and $F_{\rm 24 \mu m} \ge 0.75$ mJy MIPS catalogs with the
IRAC Shallow Survey catalog. For example, 2/3 of the compact
20 cm radio sources (excluding the OIRSs) had a 3.6 $\mu$m source counterpart centered
within $0.65\arcsec$. Since the great majority of these matches are real we took
$0.65\arcsec$ to represent the $1\sigma$ relative positional accuracy between the
radio and IRAC 3.6 $\mu$m catalogs. A similar value ($0.71\arcsec$) was found for
the $1\sigma$ relative positional accuracy between the $F_{\rm 24 \mu m} \ge 0.75$ mJy MIPS 
sources and 3.6 $\mu$m counterparts in the Shallow Survey.  The match radius of $2\arcsec$,
while seemingly small, represents essentially a $3\sigma$ criterion, i.e., we would expect
$\approx1$ radio/IRAC mismatch out of the full 377 compact radio source sample. 
As a consequence, OIMSs and OIRSs can be said to lack IRAC counterparts at the $3\sigma$
level if no 3.6 $\mu$m sources are within $2\arcsec$, and at the $5\sigma$
level if there are none within $3.3\arcsec$.

The original count of OIRSs in Higdon05 was thirty-six. However, we are no longer
confident that \#362 is a clear-cut OIRS. As a result, we will not include 
it in the following analysis. Of the remaining thirty-five OIRSs within the 
IRAC Shallow Survey, nineteen (54$\%$) were unambiguously detected at 3.6 $\mu$m as
shown in Figure 1. Even when a potential optical counterpart exists within
$\approx2-3\arcsec$ (e.g., \#97 \& 410) the compact radio source sits squarely on an
IRAC source. Of these, six (17$\%$) were detected 
in three IRAC bands to $\ge 3 \sigma$ (OIRS $\#$79, 176, 208, 245, 
346, and 349), and two ($\#$176 and 349) in all four.
The median 3.6 $\mu$m and 4.5 $\mu$m flux densities for
these nineteen objects are 18.3 $\mu$Jy and 22.5 $\mu$Jy, respectively.
Sixteen OIRSs were found to lack IRAC counterparts within $2\arcsec$ and
are shown in Figure 2. Note that in four instances ($\#$9, 305, 430, and 441) possible
IRAC counterparts are present roughly $3-5\arcsec$ away.  OIRSs $\#$9, 305, and
441 are particularly suspicious as their candidate IRAC matches are also optically 
``invisible'', i.e., like the 19 OIRSs with clear IRAC counterparts. 
The coordinate mismatches in these instances however equal or exceed the 
radio/IRAC relative positional accuracy at the $5\sigma$ level.  Moreover, the expected 
number of OIRSs in Figure 2 with an optically ``invisible'' IRAC source $3-5\arcsec$ away 
due to chance is $\approx3$, given the estimated density of 3.6 $\mu$m sources with no optical
counterparts (3.3 arcmin$^{-2}$) in the Shallow Survey.
We conclude that these four OIRSs are unlikely to be physically associated with the
relatively nearby IRAC sources.  However, even if we allow these 
four matches, it remains true that a significant fraction (at least 12/35 
or 34$\%$) of the OIRSs lack clear 3.6$\mu$m counterparts in the IRAC Shallow Survey.

It should also be noted that while the radio source morphologies in Figures 1 and 2 are 
for the most part quite simple, extended radio emission may have
been lost due to a lack of short spacings (though the 4.5-hours spent at each pointing
center in the A-array observations should reliably represent emission up to $12\arcsec$
scale-sizes) or $(1+z)^{-4}$ surface brightness dimming that {\em might} otherwise
have connected an OIRS to an optical or infrared source. Observations in more compact array
configurations will be needed to resolve this issue. With these caveats in mind,
stack averages of the 3.6, 4.5, 5.8, and 8.0 $\mu$m images at the radio source positions 
led to flux density upper-limits of 1.5, 2.0, 10.1, and 11.0 $\mu$Jy ($3 \sigma$) for the 
sixteen OIRSs with no apparent IRAC counterparts.

OIMS $\#$11 was not included in this study due to its uncertain redshift.\footnotemark[4]
\footnotetext[4]{The S/N for this source's IRS spectrum is low. While
Houck05 gives $z_{\rm spec}=0.70 \pm 0.24$, a redshift of $\approx 2.5$ is just
as feasible using the same galaxy template.} All of the sixteen remaining OIMSs
were detected by IRAC, although OIMSs $\#$17 was only observed at 3.6 $\mu$m and 5.8 $\mu$m.
These are shown in Figure 3 and their flux densities or limits are listed in Table 2. 
The median 3.6 $\mu$m and 4.5 $\mu$m flux densities for the fifteen OIMSs 
observed in all four IRAC bands 
are 17.0 $\mu$Jy and 26.1 $\mu$Jy, i.e., quite similar to that of the nineteen
IRAC detected OIRSs.  However, a much higher fraction of these OIMSs (12/15 or 80$\%$) 
were detected in all four IRAC bands. The OIMSs are also significantly 
brighter at 8.0 $\mu$m than the OIRSs: the median OIMS $F_{\rm 8.0 \mu m}$ is 
136 $\mu$Jy, while the median $F_{\rm 8.0 \mu m}$ for the nine OIRSs
detected in this band is 37 $\mu$Jy.

The OIMSs were chosen from the MIPS 24 $\mu$m catalog to satisfy
$F_{\rm 24 \mu m}$ $>$ 0.75 mJy, and the sources in Table 2 have 
a median $F_{\rm 24 \mu m}$ of 1.1 mJy.  By contrast, only four OIRSs 
were detected at 24 $\mu$m at  $5 \sigma$ or higher (OIRS $\#$97, 176, 245, 
and 363), with a mean $F_{\rm 24 \mu m}$ of 0.38 mJy. Two additional 
OIRSs were detected at the $3 \sigma$ level (OIRS $\#$208 and 
349), with a mean $F_{\rm 24 \mu m}$ of 0.24 mJy. 
We stack-averaged MIPS 24 $\mu$m sub-images 
centered on the eleven non-detected OIRSs in an attempt to recover faint 
emission from these sources. No significant signal was found, allowing us to 
set a flux density upper-limit of 67 $\mu$Jy ($3 \sigma$) at 24 $\mu$m for 
these sources as a group. No OIRSs or OIMSs
were detected with MIPS at 70 $\mu$m or 160 $\mu$m, placing $3 \sigma$ 
upper-limits of 24 mJy and 60 mJy at these two wavelengths.

The IRAC and MIPS flux densities (or $3 \sigma$ limits) for the two
source populations are listed in Tables 1 and 2. To these have 
been added flux densities (or limits) in $B_W$, $R$ 
and $I$ from the NDWFS catalog\footnotemark[5], 
and at 20 cm from either Higdon05 for the OIRSs or de Vries et al. (2002) for the OIMSs. 
\footnotetext[5]{If not already published, new $3 \sigma$ limits 
were calculated using $m_{\circ} - 
2.5~log(3~\sigma_{\rm pix}~\sqrt{\rm N_{\rm pix}})$, 
where $\sigma_{\rm pix}$ is the rms per pixel,  N$_{\rm pix}$ is the number of pixels within
the 4$\arcsec$ diameter aperture, and $m_{\circ}$ is the 
appropriate zero-point constant.}

\section{Properties Of The Optically Invisible Populations}

       \subsection{Photometric Redshifts}

Photometric redshifts ($z_{\rm phot}$) were calculated
using a set of galaxy templates corresponding to starburst (M~82, from
Xu et al. 2001), composite starburst plus AGN (Arp~220, from Devriendt
et al. 1999, modified using IRS spectra to better represent its polycyclic
aromatic hydrocarbon (PAH) emission
and silicate absorption), AGN dominated (I Zw 1, Xu et al. 2001), 
and elliptical (Silva et al. 1998) galaxies. For each OIRS, the templates 
were redshifted from $z = 0.1-6.0$ in steps of 0.05, re-gridded and scaled to
the observed flux densities. Redshifts were determined by minimizing the 
reduced $\chi^2$, defined as
\begin{equation}
{\rm
\chi^{2}_{\rm z} = N_{\rm DOF}^{-1} ~\sum \frac {(O_{\lambda} - 
\alpha_{\rm z}T_{(1+z)\lambda})^{2}}{\sigma_{\lambda}^2},
}
\end{equation}
where O$_{\lambda}$ and T$_{\rm (1+z)\lambda}$ are the observed and template
flux densities, N$_{\rm DOF}$ are the degrees of freedom,
and $\alpha_{\rm z}$ is a scaling constant calculated at each value of z. 
The optical and far-infrared upper-limits were included in the $\chi^{2}_{\rm z}$
calculation only when they were exceeded by the template flux density (i.e., when
$\alpha_{\rm z}$T$_{(1+z)\lambda} >$ O$_{\lambda}$). When this occurred O$_{\lambda}$
was set to the respective limit and $\chi^{2}_{\rm z}$ was calculated as above. 
Formal uncertainties in $z_{\rm phot}$ were estimated by numerically evaluating
\begin{equation}
{ \rm
 \sigma_{z}^{2} = \frac{2.0}{\partial^2 \chi^{2}_{\rm z}/\partial z^{2} }.
}
\end{equation}
Additional extinction corresponding to A$_{\rm V}$ = 1-2 (assuming a 
screen geometry and the Galactic reddening law in Mathis 1990) was typically 
required to satisfy the optical constraints.\footnotemark[6]
\footnotetext[6]{Estimates of the intrinsic 
extinction in the template galaxies can vary considerably depending on
angular resolution, precise location in the source, and method used.  Typical values
of A$_V$ are 80 magnitudes for Arp~220 (Devriendt et al. 1999), 10 for the
central region of I~Zw~1 (Eckart et al. 1994), to 5 for M~82 (Lester et al. 1990; Telesco
et al. 1991). The elliptical models of Silva et al. have essentially zero 
extinction.} This was applied to the template in Equation 1 prior to redshifting.

As a test, we determined photometric redshifts for the four OIMSs that Houck05 
fit with either an Arp~220 or NGC~7714 spectral template 
(OIMS $\#$2, 7, 12, and 14 in Table 2). The 
minimum-$\chi^{2}$ fits are shown in Figure 4 along with the 
corresponding photometric redshifts. We find good agreement 
between  the photometric and spectroscopic redshifts of all four, with 
$\vert z_{\rm phot} - z_{\rm spec}\vert$ = 0.1, 0.1, 0.1 and 0.3, 
respectively. The Arp~220 template with an additional $\approx 1.5$ 
magnitudes of visual extinction provided the best fits for all
four, including OIMS $\#$2, whose IRS spectrum was classified as
a starburst (NGC~7714) by Houck05. Starburst templates gave similar 
$z_{\rm phot}$ for this source, but produced too little emission at 
8.0 $\mu$m and 24 $\mu$m, implying the presence of an additional dust
component in the rest-frame mid-IR. The fact that OIMSs \#12 \& 14 are 
significantly brighter than the Arp~220 template in the 24 $\mu$m band may reflect 
stronger PAH emission in these sources. Again, similar photometric 
redshifts were obtained using the M~82 template and down-weighting the 24 $\mu$m
measurements, however only the Arp~220 template could match the observed 
$F_{\rm 24 \mu m}/F_{\rm 8.0 \mu m}$.  The remaining
OIMSs in Table 2 either lacked a sufficient number of data points for template fitting
(e.g., $\#$5, 10, and 17)  or possessed power-law SEDs, for which unambiguous 
redshift determinations were not possible.

Photometric redshifts were derived for six
OIRSs: $\#$97, 176, 208, 245, 349, and 363. Each was
detected in at least three IRAC and MIPS bands to $\ge 3 \sigma$.
The minimum-$\chi^{2}$ fits are shown in Figure 5, along with the
best matching template and derived $z_{\rm phot}$. The results are summarized 
in Table 3. In each case we are fitting the inflection point in the
spectrum where the dominant emission source changes from evolved 
stars to warm dust. For OIRS $\#$363 the
inflection point is implied by the upper-limits in the IRAC 5.8 $\mu$m and 8.0 
$\mu$m bands, which effectively rule out a power-law between 3.6 $\mu$m
and 24 $\mu$m. The derived redshifts range from
1.0 to 4.5 with a mean of 2.2. The formal redshift uncertainty is 
$\approx0.2-0.6$, though a comparable contribution to the error budget may 
arise from our ignorance of the source's true SED. For example, OIRSs
$\#$97 and 245 would have been detected at 160 $\mu$m to $> 3 \sigma$
had the Arp~220 template accurately represented their far-infrared SEDs.
(A more in depth analysis employing a larger set of templates 
and extinction laws is not warranted at this point given the paucity of data points.) 
The six OIRSs are best represented by either the starburst M~82 
($\#$176, 208, and 349) or the starburst/AGN composite Arp~220 ($\#$97, 245, and 
363) templates. In each case the best fitting template was unambiguous. In particular,
the AGN-dominated galaxy templates make poor matches to the SEDs 
in Figure 5. Likewise, elliptical galaxy templates are ruled out by the 
detection of emission at 24 $\mu$m. This sub-sample of OIRSs detected both
with IRAC and with MIPS at 24 $\mu$m represents a population of sources with a significant
starburst component with a redshift range similar to the OIMSs.

Two other OIRSs, $\#$79 and 346, permitted rough photo-$z$ estimates.
Minimum-$\chi^{2}$ fits are shown in Figure 6. For $\#$79, the flat
IRAC spectrum could reasonably be interpreted as the rest-frame $\sim$1-3
$\mu$m region, consistent with redshifts between $1.5-2.5$. In fact, the
smallest $\chi^{2}$ is achieved with a starburst (M~82) template at
$z_{\rm phot}$= 1.6 with one magnitude of additional extinction.
An Arp~220-like SED appears to be ruled
out unless the 9.7 $\mu$m silicate absorption feature
in these sources is much deeper than the template. 
We emphasize that these redshifts are approximate,
as $z_{\rm phot} = 2.5$ also yields a small $\chi^2$.
The second OIRS in Figure 6, $\#$346, shows a fairly steep
power-law shape ($\alpha$ = -2.1, for $F_{\nu} \propto \nu^{\alpha}$) 
in the IRAC data that does not extend to the MIPS 24 $\mu$m band. If this spectral
break occurs at rest-frame wavelengths of 1-2 $\mu$m, then
redshifts exceeding 3 are implied, and a reasonable fit using
the M~82 and Arp~220 template at $z_{\rm phot} = 3.5-4.6$ are possible,
as is shown in Figure 6. Note that good
fits also result using the elliptical template for both,
since the 24 $\mu$m upper-limit provides no constraint to the
shape of the SED.
In both cases, the derived $z_{\rm phot}$ is close to that derived using 
M~82 or Arp~220. Because these redshifts are only estimates, we do
not include them in Table 3. 
 
Overall, no starburst, AGN, or composite template could be found that would 
permit a fitted $z_{\rm phot}$ of less than one for the above eight OIRSs. 
At the very least, a significant fraction of 
OIRSs are at high redshift. While galaxy templates with a sizable young 
stellar component (i.e., Arp~220 or M~82) are consistent with the data, 
early-type galaxies cannot be ruled out in the two cases where there are no 24 $\mu$m
detections (Figure 6). 

There are insufficient data to determine photometric redshifts
for the remaining OIRSs in Table 1. It is important to keep in
mind that we have derived reasonably confident $z_{\rm phot}$ for only 1/3
of the IRAC detected OIRSs. This represents only $17\%$
of the total OIRS sample from Higdon05.  We will address the issue 
of how likely these six are to be representative of the entire OIRS 
population below.

       \subsection{Spectral Energy Distributions and Emission Mechanisms}

Houck05 found that the mid-IR spectra of the OIMSs could be
fit using only four templates, all derived from local galaxies (see Table 4). 
The same templates also fit the wider spectral energy distributions examined 
here. They are: (1) a heavily obscured Seyfert~2 nucleus, represented by IRAS 
F00183-7111 and denoted {\it obsSy2}, (2) an obscured Seyfert~1 nucleus, 
represented by Mrk~231 and denoted {\it obsSy1}, (3) a starburst nucleus, 
represented by NGC~7714 and denoted {\it sb}, and (4) a combination of an 
obscured AGN and a starburst, represented by Arp~220 and denoted {\it comb}. 
The first four panels of Figure~7 show average rest-frame SEDs for the
sources in each category. The SEDs were found by shifting each source's flux 
densities (or limits) to the rest-frame, then normalizing at $\lambda_{\rm rest} 
= 1.4$~\micron.  The individual scaled SEDs were then coadded and binned.

The averaged SEDs for the {\it obsSy1} and {\it obsSy2} OIMSs both show steep
power-laws extending from $\lambda_{\rm rest} \approx 1 - 10~\micron$.  This is
consistent with a luminous and heavily obscured active
nucleus. Assuming this emission arises from dust, which is reasonable given
the high extinction deduced by Houck05, emission at $\sim$1~\micron\
implies that some of the dust approaches the
sublimation temperature of refractory dust.  For the {\it obsSy2} SED, the
power-law exponent $\alpha \approx -4.0$ ($F_{\nu}\propto
\nu^{\alpha}$) over the range $1 \la \lambda_{\rm rest}\la
4$~\micron, and there is a break in the power-law at $\lambda_{\rm
rest}\sim 5$--8~\micron.  For the {\it obsSy1} OIMSs, the slope is
shallower ($\alpha \approx -2.3$), but the power-law component extends
over a wider range, $1 \la \lambda_{\rm rest} \la 10 ~\micron$, with no
break.  The slope is still steep compared to SEDs of unobscured AGN
(e.g., Elvis et al. 1994) and even compared to the lensed $z = 3.91$
quasar APM~08279+5255 (Soifer et al.\ 2004), for which $\alpha = -2.0$
over $1\la \lambda_{\rm rest} \la 4$~\micron.  The steep SEDs suggest
a distribution of dust mass that rises rapidly with distance from the
central heating source, and the break in the {\it obsSy2} SED suggests that
the rapid rise ends at a distance corresponding to $T_{dust}\approx
400$~K.  The absence of PAH emission in the IRS spectra of these
objects is consistent with the AGN interpretation of the emission.

The {\it sb} SED rises about as steeply ($\alpha\approx -2.3$) as the
{\it obsSy1} SED in the rest-frame 1--8~\micron\ spectral range, though with only
one object the slope is poorly determined.  A 1.6~\micron\ bump can
be seen in the IRAC data, indicating that the rest-frame 1--3~\micron\
emission arises primarily from evolved stars.  The steep rise at longer
wavelengths is characteristic of dust in star formation regions.  In
contrast, the average SED for {\it comb} objects is much shallower for
$\lambda_{\rm rest}$ between 1 - 3~\micron, $\alpha\approx -1.1$, with a steeper rise from
rest 3 to 8~\micron. The lack of a prominent 1.6~\micron\ bump
suggests a power-law AGN component contributes much of the rest-frame
$1-3$ $\mu$m flux, and the steepening suggests the presence of dust
associated with the starburst, consistent with the dual starburst/AGN nature
of these objects.

The average SED for the six IRAC-detected OIRSs with photometric
redshifts is also shown in Figure~7. This SED resembles the {\it sb}
class by virtue of the clear 1.6 $\mu$m bump. However, the steep rise for
$\lambda_{\rm rest}$ between 3 - 10 $\mu$m may reflect a combination
of star formation plus a weak active nucleus. In any event,
there is no evidence of a dominant obscured active nucleus, which
suggests that a significant fraction of the luminosity is provided by
star formation. This is in marked contrast with the OIMSs, which are
powered in most cases by dusty AGN.  We again wish to stress that the 
averaged OIRS SED was created using only the six sources with photometric
redshifts and may not be representative of the remaining thirteen
IRAC detected OIRSs in Table~1, much less the parent population of
OIRSs in Higdon05.

Additional insight into the dominant power source for
OIMSs and IRAC detected OIRSs can be gathered from their positions
in an IRAC two-color diagram. Figure 8 shows that OIMSs and OIRSs tend to 
occupy distinct regions. Most OIMSs define
a fairly narrow distribution extending from relatively flat observer-frame
SEDs ($F_{\rm 8.0 \mu m}/F_{\rm 4.5 \mu m} = 
F_{\rm 5.8 \mu m}/F_{\rm 3.6 \mu m} \approx 1$) in the lower-left 
to steep power-law SEDs ($F_{\rm 8.0 \mu m}/F_{\rm 4.5 \mu m} =
F_{\rm 5.8 \mu m}/F_{\rm 3.6 \mu m} \approx 10$) in the 
upper-right. Nearly all of the {\it obsSy1} and {\it obsSy2} OIMSs 
lie above $F_{\rm 8.0 \mu m}/F_{\rm 4.5 \mu m} = 3$, in the portion of the
diagram dominated by quasars and obscured AGN (Lacy et al. 2004). This is 
consistent with both their IRS spectra (which show deep 9.7 $\mu$m silicate
absorption and no PAHs) and their steep power-law SEDs in Figure 7.
The {\it comb} and {\it sb} OIMSs are situated at $F_{\rm 8.0 \mu m}/F_{\rm 4.5 \mu m} 
\la 3$ and $F_{\rm 5.8 \mu m}/F_{\rm 3.6 \mu m} \la 2$, where
starburst systems dominate, also in agreement with their IRS spectra.

The IRAC detected OIRSs are shown in the lower panel of Figure 8.
The six OIRSs with photometric redshifts have positions that are for the
most part consistent with the dusty M~82 track for $z \approx 1-3$, 
in agreement with their $z_{\rm phot}$ and fitted spectral templates.
The remainder of this population are also clustered in the lower-left 
of the figure, i.e., $F_{\rm 8.0 \mu m}/F_{\rm 4.5 \mu m} < 3$ and $F_{\rm 5.8 
\mu m}/F_{\rm 3.6 \mu m} < 4$. Given the uncertainties, their IRAC colors are
equally consistent with elliptical galaxies at $z > 2$ or moderately obscured 
($A_{\rm V} \approx 2.5$) $z > 1$ starbursts. 
However, all possess radio-loud AGN by virtue of their negative {\it q}-values
and so cannot be pure starbursts.

Figure 8 highlights the fact that the IRAC detected OIRSs as a group have
much flatter mid-IR SEDs than the OIMSs. The latter possess mean 
$F_{24 \mu m}/F_{3.6 \mu m}$ ratios of 170, 65, 52, and 44
for the {\it obsSy2}, {\it obsSy1}, {\it sb}, and {\it comb} classes respectively. By
contrast, the thirteen IRAC detected OIRSs without $z_{\rm phot}$ 
(i.e., the filled circles in Figure 8) possess a mean
$F_{24 \mu m}/F_{3.6 \mu m} < 11$.  Given their 3.6~\micron\ flux
densities, all thirteen should have been detected at 24~\micron\ had
their SEDs been steep power-laws like the {\it obsSy1} and {\it obsSy2} 
OIMSs.\footnotemark[7] \footnotetext[7]{ For example, OIRS~\#232, with the smallest 
$F_{3.6 \mu m}= 8.1\pm1.7$~$\mu$Jy, would have yielded 24~\micron\ flux densities 
between 0.5 and 1.4~mJy, easily detectable by the MIPS survey.}
Even if their SEDs resembled the {\it sb} or {\it comb} OIMSs, most would have been
detected at 24 $\mu$m.  This is consistent with emission from stellar photospheres and 
AGN heated dust being at least {\it comparable} in most of the IRAC detected OIRSs.

       \subsection{Luminosities and Stellar Masses}

Rest-frame K-band luminosities, defined as  L$_{\rm K} = 
\Delta \nu {\it L}_{\nu}(2.2 \mu m)$, with $\Delta \nu$ corresponding to the 
Johnson K-band filter FWHM ($2.1 \times 10^{13}$ Hz; 
Tokunaga et al. 2002) were calculated for the OIMSs and the six OIRSs with photometric 
redshifts. These are listed in Tables 3 and 4 in units of 
the sun's K-band luminosity (L$_{\odot,\rm K}$ = 8.7 $\times$ 10$^{24}$ W;
Bessel 1979).  With this definition, Arp~220 and IRAS F00183-7111 would have Johnson 
K-band luminosities of $1.3 \times 10^{11}$ L$_{\odot,\rm K}$
and $9.1 \times 10^{11}$ L$_{\odot,\rm K}$ respectively.  Monochromatic
rest-frame luminosities at 5 $\mu$m, defined as L$_{\rm 5 \mu m} = \nu L_{\nu}(\rm 5 \mu m)$,
were also calculated for these sources. This quantity provides a measure of the emission from
hot ($400 - 1500 ~K$) dust free of any contribution from
evolved stars and which avoids the (rest-frame) $9.7 \mu m$ silicate feature. 
These are also given in Tables 3 and 4. Uncertainties in redshift constitute the 
dominant source of error in the derived luminosities.

The OIMSs are highly luminous in the rest-frame K-band, with a median L$_{\rm K}$ of
$1.6 \times 10^{12}$ L$_{\odot,\rm K}$ and more than an order of magnitude
spread. The {\it obsSy2} OIMSs are the most luminous, with L$_{\rm K}$
$\approx 10^{13}$ L$_{\odot,\rm K}$ on average. These are followed by the
{\it obsSy1} and {\it comb} OIMS, both with average L$_{\rm K}$
$\approx 10^{12}$ L$_{\odot,\rm K}$. The {\it obsSy2} and {\it comb} OIMSs are 
both on average an order of magnitude more luminous than their local templates, 
IRAS F00183-7111 and Arp~220. Their rest-frame monochromatic 5 $\mu$m luminosities 
are also extremely large, with median L$_{\rm 5 \mu m}$  = $1.0 \times 10^{39}$ W and
$1.2 \times 10^{38}$ W for the {\it obsSy2} and {\it obsSy1}
OIMSs respectively. 

The six IRAC detected OIRSs with photometric redshifts  show a very wide 
range in L$_{\rm K}$,  led by $\#$176 with $2.8 \times 10^{13}$ L$_{\odot,\rm K}$. 
This is roughly fifty times more luminous than the five other OIRSs and nearly
twice as large as the most luminous OIMS.
The three OIRSs that were best fit with a {\it comb} template  
have a mean  L$_{\rm K}$ of 5 $\times$ 10$^{11}$ L$_{\odot,\rm K}$.
This is within a factor of two of the average for the three {\it comb} OIMSs
(OIMS $\#$7, 12 and 14).
Accurate 8-1000 $\mu$m luminosities (L$_{\rm IR}$) for the OIMSs and
OIRSs are not possible since none were detected at 70 $\mu$m or
160 $\mu$m. However, the fact that median L$_{\rm K}$ for the OIRSs
and {\it comb/sb} OIMSs are similar suggests that they have
comparable total luminosities, and that at least some of the IRAC detected
OIRSs are hyper-luminous.

The 1.6 $\mu$m bump apparent in the averaged OIRS SED in Figure 7
points to a significant evolved stellar component. For these
objects we estimated stellar masses by multiplying their L$_{\rm K}$
by near-infrared mass to light ratios derived from burst models
in Bell \& de Jong (2001). From their Figure 1, we adopted
M$_{*}$/L$_{\rm K}$ = 0.3, which corresponds to their bluest models 
($B-R$ = 0.6). The mass to light ratio increases fairly slowly with
color, as can be seen from the fact that M$_{*}$/L$_{\rm K}$ = 0.5 at $B-R$ = 1.0.
We find a wide range in derived stellar masses, again led by OIRS $\#$176
with M$_{*} \approx 10^{13}$ M$_{\odot}$. The other OIRSs have masses in the range
$0.8 -3.4 \times 10^{11}$ M$_{\odot}$, with a mean of $\approx 2 \times 10^{11}$
M$_{\odot}$. Very massive galaxies are indicated even if warm dust 
makes a 50$\%$ contribution to the rest-frame 2.2 $\mu$m emission.

Despite the lack of redshifts for the remaining IRAC detected OIRSs, we can
set useful limits to their stellar masses if we assume they
lie at $z \ge 1$ and that the (putative) AGN makes an insignificant contribution 
to the IRAC 3.6 $\mu$m and 4.5 $\mu$m bands. The first assumption is reasonable 
given their faint optical limits. The second follows from the data in Table 1, where 
only OIRSs \#49, 79, 232, \& 410 have mid-IR to radio SEDs that might allow 
significant synchrotron emission in the mid-IR.  
For $z = 1$, where the IRAC 4.5 $\mu$m band measures
rest-frame 2.2 $\mu$m emission, the average $F_{\rm 4.5 \mu m}$ (18.0
$\mu$Jy) implies a K-band luminosity of $1.5 \times 10^{11}$ L$_{\odot, \rm K}$. 
For an early-type M$_*$/L$_{\rm K}$ ratio, we derive a typical stellar mass of 
$\approx 5 \times 10^{10}$ M$_{\odot}$
for this sub-population. So long as there is minimal contribution from an AGN to
the 4.5 $\mu$m emission, the IRAC detected OIRSs without redshifts
also appear to be massive galaxies.

For the majority of OIMSs, the rest-frame near-infrared is dominated by
non-stellar emission, and the upper-limits to their stellar masses are not
very informative.  However, the four OIMSs with {\it comb} or {\it sb} IRS spectra 
(\#2, 7, 12, and 14) show evidence of a weak 1.6 $\mu$m bump in Figures 1 and 4. 
Using the same starburst M$_{*}$/L$_{\rm K}$ of 0.3, their rest-frame
L$_{\rm K}$ imply {\it maximum} M$_*$ between $2-4 \times 10^{11}$ M$_{\odot}$. 
These objects also appear to be very massive even if the stellar component makes 
only a $\sim20\%$ contribution to the rest-frame near-infrared.

Rest-frame 20cm luminosity densities were estimated for the OIMSs and
the six IRAC detected OIRSs with $z_{\rm phot}$ using
\begin{equation}
{\rm
L_{\rm 20 cm} = 1.2 \times 10^{20}~D_{\rm L}^{2}~(1+z)^{\alpha-1}~F_{\nu}(\rm 20 cm)~~~W~Hz^{-1},
}
\end{equation}
where D$_{\rm L}$ is the luminosity distance in Mpc, $\alpha$ is
the radio spectral index ($F_{\nu} \propto \nu^{-\alpha}$), taken to be 0.65, 
and $F_{\nu}(20 cm)$ is the observed 20 cm flux density (or $5 \sigma$ upper-limit)
in Jy.  The latter are taken from Higdon05 and de Vries et al. (2002).
Seventy per cent of the 
OIMSs were  not detected in the radio, implying L$_{\rm 20 cm}$ upper-limits of 
$\sim 5 \times 10^{24}$ W Hz$^{-1}$ ($5 \sigma$). Of the remainder, OIMS $\#$13 in Table 4 is 
noteworthy for possessing the largest radio luminosity density among both OIRS and OIMS,
with L$_{\rm 20cm} = 1.0 \times 10^{26}$ W Hz$^{-1}$. It represents
the sole OIMS that is squarely in the ``powerful'' radio galaxy class (i.e., L$_{\rm 20cm}$
$\ge 10^{25}$ W Hz$^{-1}$; McCarthy 1994). A wide range of values is found for the 
OIRSs with photometric redshifts, with L$_{\rm 20 cm}$ between 0.2-2.9 $\times$ 10$^{25}$ 
W Hz$^{-1}$ and a median of $4 \times 10^{24}$ W Hz$^{-1}$. 

The six IRAC detected OIRSs show signs of large evolved stellar populations
by virtue of the 1.6 $\mu$m bump apparent in Figure 7.
If we assume that their radio emission arises primarily from supernova remnants, we
can estimate their current star formation rates with their L$_{\rm 20cm}$. Excluding
the radio-loud OIRS \#97 ({\it q}$  ~= -0.24$), we derive a median SFR of
2200 M$_{\odot}$ yr$^{-1}$ after Bell (2003). For the single {\it sb}-like
OIMS the L$_{\rm 20cm}$ upper-limit implies SFR $\la 2000$ M$_{\odot}$ yr$^{-1}$.  
Assuming a minimal contribution
to L$_{\rm 20cm}$ from an active nucleus, we find these objects to possess
SFRs comparable to the highest values derived for local ultra-luminous
infrared galaxies (ULIRGs) using both X-ray and far-infrared luminosities 
(e.g., Persic et al. 2004).

\subsection{The Near Environments of OIMSs and OIRSs}

Most of the IRAC detected OIRSs appear to possess nearby companion candidates, whereas
few of the OIMSs do. Figure 3 shows that the OIMSs are typically unresolved by IRAC at 
3.6 $\mu$m, with only OIMS \#3 and 6 showing hints of structure. By contrast,
twelve of the nineteen IRAC detected OIRSs ($63 \pm 18\%$) in Figure 1 are clearly 
not simple point-sources. Examples include OIRS \#79, 97, and 363. 
While gravitational lensing by foreground clusters can not be ruled out,
the most straightforward interpretation is that IRAC is detecting
several closely spaced galaxies. This is consistent with the majority
of the IRAC detected OIRSs having companions within $\approx 50 ~kpc$.

The clustering properties of these systems are beyond the scope of this
paper. It is worth noting, however, that neither OIMSs nor OIRSs
appear to inhabit particularly crowded environments like the cores of rich clusters. 
For example, using the sixteen IRAC subimages of OIMSs in Figure 3 we find an
average of $1.31 \pm 0.29$ sources per $20\arcsec$ wide field (not including the OIMSs). 
The average number of sources in the fields centered on OIRSs is $1.74 \pm 0.30$,
again not including the OIRSs. As a baseline figure, we would have expected 
$1.34 \pm 0.01$ sources in a random $20\arcsec$ sized region given the total number 
of sources ($\approx370,000$ at 3.6 $\mu$m) detected within the 8.5 deg$^2$ IRAC 
Shallow Survey (Eisenhardt et al. 2004).\footnotemark[8] 
\footnotetext[8]{We used the Student t-Test to 
evaluate the likelihood that the IRAC detected OIRSs have a significantly higher
number of companion candidates within the 20$''$ wide fields relative to the OIMSs.
For the sources visible in the two sets of IRAC sub-images, we calculated t = 1.43. 
For thirty-two degrees of freedom and a 95\% confidence level, t would have had to 
exceed 2.04 for a significant difference between the OIMSs and OIRSs near environments 
to be real. We thus find no measurable difference in the apparent density on this scale.}

This is not to suggest that OIMSs and the IRAC detected OIRSs do not inhabit clusters.
Recent work indicates that radio galaxies and quasars show a wide variety in
local density up to at least $z \approx 1.5$, from the cores of rich clusters to
field-like environments (e.g., Barr et al. 2004; Best et al. 2003).
Further, the radio galaxies and radio-loud quasars that are associated with rich
centrally concentrated clusters are as often as not situated outside the actual
core. Given these trends and the rather small spatial dimensions of the IRAC
subimages (area $\approx$ 0.04 Mpc$^2$) it is perhaps not surprising that we do
not see an excessive number of companion galaxies surrounding the OIMSs or OIRSs.
An investigation of the environments of OIMSs and OIRSs on scales of several Mpc$^2$
will be presented in a future paper.

\section{Discussion}

\subsection{OIMSs as Buried QSOs}

Houck05 concluded that both the {\it obsSy1} and {\it obsSy2} OIMSs are
heavily obscured and hyper-luminous, with L$_{\rm IR}$ estimated to be 
$\approx 10^{13} ~L_{\odot}$. Their IRS spectral shapes, and in particular,
the absence of PAHs, implied that their primary energy source was an active
nucleus rather than star formation. 
We have shown that the {\it obsSy1} and {\it obsSy2} OIMSs also possess steep
power-law SEDs over the rest-frame $1 - 10 ~\mu$m region and
extremely high rest-frame monochromatic 5 $\mu$m luminosities (L$_{\rm 5 \mu m} 
= 10^{38} - 10^{39.2}$ W). The latter originates in the hottest dust component
and necessarily implies a powerful active nucleus.
To place these  luminosities in context, we calculated rest-frame 
$\nu L_{\nu}(5 \mu m)$ for the sample of 42 3CRR radio galaxies and quasars observed with
{\it Spitzer} by Ogle et al. (2006).\footnotemark[9]
\footnotetext[9]{Rest-frame $\nu L_{\nu}(5 \mu m)$ were extrapolated for the
quasars and radio galaxies in Table 1 of Ogle et al. using
the listed rest-frame $\nu L_{\nu}(15 \mu m)$ and the 7 to 15 $\mu$m
spectral indices.}  For the sources classified as {\it mid-IR
luminous}, which consist of quasars and narrow-line radio galaxies (NLRGs),
L$_{\rm 5 \mu m} = 10^{37.4} - 10^{39.3}$ ~W. 
The quasar-like L$_{\rm 5 \mu m}$ of the {\it obsSy1} and {\it obsSy2} OIMSs,
together with their steep power-law SEDs and IRS spectra provide compelling
evidence that they are driven by powerful AGN accreting near the Eddington limit.
These sources may be considered to be obscured or ``buried'' quasi-stellar objects (QSOs).

Final proof that the {\it obsSy1} and {\it obsSy2} OIMSs host heavily
obscured QSOs would be the detection of luminous hard (i.e., $> 2$ ~keV) X-ray  emission. 
These photons originate very near the super-massive black hole
and are a reliable indicator of powerful AGN activity. They are also
fairly insensitive to obscuration, at least up to
N$_{\rm H} = 1.5 \times 10^{24}$ cm$^{-2}$ where the optical depth for
Compton scattering reaches unity.  Of the OIMSs discussed in this paper, only
$\#$13 (not coincidently, the most radio-loud OIMS) was detected above the 99$\%$ confidence level
in the ``hard'' ($F_{\rm 2 - 7 ~keV} = (1.6 \pm 0.8) \times 10^{-14}$ erg s$^{-1}$ cm$^{-2}$) 
and  ``full'' ($F_{\rm 0.5 - 7 ~keV} = (2.3 \pm 1.4) \times
10^{-14}$ erg s$^{-1}$ cm$^{-2}$) bands in the 2 ks {\it Chandra} 
survey of the Bo\"{o}tes field by Kenter et al. (2005). This corresponds to a hard X-ray 
luminosity L$_{\rm 2 - 7 ~keV} = (5 \pm 2)\times 10^{37}$ W, with no correction 
for intrinsic extinction.\footnotemark[10] 
\footnotetext[10]{X-ray luminosities were calculated from
L$_{\rm 2-7 ~keV} = 4 \pi D^{2}_{\rm L}~F_{\rm 2-7 ~keV} (1 + z)^{\Gamma - 2}$, where
D$_{\rm L}$ is the luminosity distance and $\Gamma$ is the power-law index, taken to be 1.8.} 
For the other eight {\it obsSy1} and {\it obsSy2} OIMSs we obtain upper-limits of
L$_{\rm 2-7 ~keV} \la 10^{37.5}$ W, again with no extinction correction.
Powerful AGN in the local universe typically possess L$_{\rm 2-7 ~keV} \approx 10^{35}-10^{37}$ 
W. So while only OIMS $\#$13 can be conclusively shown to be a ``buried''
QSO based on its hard X-ray emission, the other {\it obsSy1} and {\it obsSy2} OIMSs are 
still consistent with this interpretation.
Deeper observations (or a stacking analysis) to detect hard X-rays with {\it Chandra}
should be carried out, as well as searches for high-ionization infrared lines such
as [S~IV] $\lambda 10.51~\mu m$ and [Ne~V] $\lambda 24.31~\mu m$. These were not evident 
in the IRS spectra shown in Houck05, though this can be attributed to the low spectral resolution
coupled with the relative faintness ($F_{\rm 24 \mu m} \approx 1$ mJy) of
the sources. 

\subsection{The OIRSs Detected by IRAC}
The IRAC detected OIRSs fall into two groups depending on whether or not they
were also detected with MIPS at 24 $\mu$m. Those that are detected at 24 $\mu$m
appear to be massive and dusty $z \approx 2$ galaxies that are primarily driven by
star formation. Those not detected at 24 $\mu$m (by far the majority) possess
radio-loud AGN as evidenced by their negative {\it q} upper-limits. They also
appear to be massive, with $M_{*} \approx 10^{11} M_{\odot}$ if they are at $z > 1$
as their faint optical limits imply. Their individual
$F_{\rm 24 \mu m}/F_{\rm 3.6 \mu m}$ limits are consistent with mid-IR SEDs 
in which stellar emission is at least comparable to that from dust
close to the central engine.

Figure 9 shows the ratio $F_{\rm 24 \mu m}/F_{\rm 3.6 \mu m}$ as a function of redshift
for five representative objects known to contain powerful AGN: the composite
{\it obsSy2} OIMS from Figure 7, the NLRG 3C459, the quasar 3C273, the Low Excitation 
Radio Galaxy (LERG) 3C293, and the ``red and dead'' radio galaxy LBDS~53W091 (Stern 
et al. 2006). Except for the OIMS and LBDS~53W091 the SEDs were constructed using 
data from NED. This figure also illustrates the extremely steep $\lambda_{\rm rest} =
1-10$ $\mu$m SED of the {\it obsSy2} OIMSs relative to a fairly typical
NLRG and quasar. The thick horizontal line
in the figure represents the mean $F_{\rm 24 \mu m}/F_{\rm 3.6 \mu m}$ upper-limit  
for the IRAC detected OIRSs ($F_{\rm 24 \mu m}/F_{\rm 3.6 \mu m} < 3.9$). This
was determined by dividing the stack-averaged 24 $\mu$m 
upper-limit (67 $\mu$Jy, $3 \sigma$) by the mean $F_{\rm 3.6 \mu m}$ for the twelve OIRSs
not detected at 24 $\mu$m. The upper/lower bounds to these limits, defined by
dividing the stack-averaged 24 $\mu$m limit by the minimum/maximum $F_{\rm 3.6 \mu m}$ values 
are also shown as two thin horizontal lines. Taken as a group, these OIRSs have spectral shapes that
are inconsistent with {\it obsSy2} OIMSs,  typical NLRGs, and quasars at essentially all redshifts. 
They are, however, consistent with LERGs and the ``red and dead'' radio source LBDS~53W091 
for $z > 1$.

Of the two, LERGs are currently the best studied at infrared 
wavelengths. LERGs are identified by their nuclear spectra which show
extremely weak high-excitation optical lines like [O~III] $\lambda$5007 \AA
~(Laing et al. 1994), and are nearly always hosted by massive
ellipticals. Most FRI radio galaxies are LERGs, though a substantial number of FRIIs 
are as well. LERGs display a wide range of radio luminosities, with L$_{\rm 20 cm}$ $\approx$ 
10$^{23}$ W~Hz$^{-1}$ to $>10^{26}$ W~Hz$^{-1}$, showing that LERGs can be 
``powerful'' radio galaxies. Their flat
rest-frame near to mid-IR SEDs have been attributed to the absence of massive obscuring
tori like those inferred in NLRGs and BLRGs. More direct evidence comes from X-ray spectra
of LERG nuclei, which (with few exceptions) lack an absorbed soft component
(Evans et al. 2006), and radio VLBI observations showing polarized emission
on parsec scales (Kharb et al. 2005).\footnotemark[11] Siebenmorgen et al. (2004) used 
simple dust models to fit the optical/infrared
SEDs of a number of LERGs in the 3CRR (Laing et al. 1983). They concluded that compared 
with NLRGs and BLRGs, LERGs possess lower AGN luminosities, more extensive
dust distributions (i.e., much less dust at small radii), and lower overall 
obscuration (A$_{\rm V} \approx 2$). \footnotetext[11]{This has led to suggestions 
that the accretion process in 
LERGs is fundamentally different from NLRGs and BLRGs (e.g., Hardcastle et al. 2006).
Accordingly, super-massive black-holes in NLRGs and BLRGs accrete material from a dusty 
molecular torus while those in LERGs accrete hot gas directly from the intergalactic medium 
through Bondi accretion (Allen et al. 2006). For our purposes, this is less important 
in understanding their infrared SEDs than the apparent lack of a massive obscuring torus
extending close to the central engine.} 

The small $F_{\rm 24 \mu m}/F_{\rm 3.6 \mu m}$ limits for the OIRSs
not detected at 24 $\mu$m are consistent with a population of high-$z$ LERGs, i.e., 
massive elliptical galaxies hosting radio-loud AGN with relatively small amounts 
of dust close to the central engine. 
However, the data are also consistent with a population of radio-loud $z > 1$ 
ellipticals with active nuclei weak enough to be ``drowned out'' by stellar emission,
even if only temporarily, due to a drop in the mass accretion rate.

Whether the OIRSs discussed in this section are LERGs or not, their
radio luminosities will be substantially less than $10^{25}$ W Hz$^{-1}$ if their typical
redshifts are $\approx 1$. Only if $z \ga 2$ will most of these objects be
classified as ``powerful'' radio galaxies. Future observations to better constrain 
their rest-frame infrared SEDs and determine
accurate redshifts will be required to help resolve this issue.

It may at first seem puzzling that so many of the IRAC detected OIRSs resemble LERGs and
``red and dead'' radio galaxies. One may ask where are the optically ``invisible'' mid-IR 
luminous NLRGs? Part of the answer must lie in the way the OIRS sample was originally
defined, namely, by requiring a compact or unresolved radio morphology in the 
A-array 20 cm images. The original intent of the VLA survey was to detect
optically obscured $z > 1$ {\it starburst} galaxies through their radio emission. 
Such sources were expected to be point-like in the 20 cm A-array images.
As a result, extended radio structures (i.e., jets and lobes) were excluded from 
consideration from the outset since they (1) would identify AGN dominated
sources, and (2) would have also proven difficult to confidently associate with an optical
counterpart given the high source density in the NDWFS $B_W$, $R$, and $I$-band images. 
As a consequence, it is possible that we  preferentially selected beamed radio sources, i.e., 
sources resembling BL Lac objects. It has been suggested that FRI radio galaxies 
are the parent population for BL Lac objects. The fact that LERGs comprise most 
of the local FRI allows a simple explanation, namely, that most of the IRAC selected OIRSs 
are in fact a population of high-z BL Lac objects.

	\subsection{The OIRSs Not Detected by IRAC}

The fact that nearly half (16/35) of the OIRSs appear to lack IRAC counterparts
(see Figure 2) is a significant result, especially in light of the small
number of spurious radio sources expected in Higdon05.\footnotemark[12]
\footnotetext[12]{There are 392 radio sources in the final catalog. The number of false radio 
sources was estimated by repeating the SExtractor runs on the final VLA images multiplied 
by -1. The number of $> 5 \sigma$ negative ``sources'' provides a robust estimate of the 
number of false positives. Only three such sources were found in the radio images, a number 
that is also consistent with random fluctuations given the number of independent synthesized beams 
in the radio images. Minimal contamination of the source catalog is indicated. We also considered
the possibility that some of the OIRSs might be hot-spots in large radio doubles ``resolved
out'' by the interferometer array. However, given their wide separation (few OIRSs are within 
2-3 minutes of each other) these would have to be gigantic radio sources, and therefore unlikely.}
By contrast, all of the OIMSs were detected by IRAC, which suggests that 
the two groups have very different SEDs and/or redshifts. We can constrain 
the average L$_{\rm K}$ of the non-IRAC detected OIRSs (and indirectly, their likely 
redshift range) with the stack-averaged $3 \sigma$ upper-limits at 3.6, 4.5, 5.8, 
and 8.0 $\mu$m listed in $\S$2. 

If the OIRSs not detected by IRAC are primarily a $z <$ 1 population, they must 
be sub-L$^*$ galaxies.
For an object at $z = 0.64$, where rest-frame 2.2 $\mu$m shifts to the center
of the IRAC 3.6 $\mu$m band, the 1.5 $\mu$Jy upper-limit would correspond to
L$_{\rm K}$ $<$ 1.2 $\times$ 10$^{10}$ L$_{\odot, \rm K}$. 
This is substantially smaller than the K-band luminosities of L$^*$ galaxies determined
by Kochanek et al. (2001), who found L$_{\rm K}^{*}$ = 0.7-1.2 $\times$ 10$^{11}$
L$_{\odot, \rm K}$ for early to late types.\footnotemark[13] \footnotetext[13]{
Kochanek et al. derived M$^*_{\rm K}$ using a complete sample of 4,353 galaxies
in the 2MASS survey. For H$_{\circ}$ = 71 km s$^{-1}$ Mpc$^{-1}$, their results
translate into M$_{\rm K}^*$ = -23.7 (late-type), -24.3 (early-type), and -24.1 (all types).}
This in turn limits the stellar mass of the non-IRAC detected OIRS hosts.
For an early-type galaxy M$_{*}$/L$_{\rm K} \approx 0.7$ (Bell \& de Jong 2001), and
the resulting M$_{*}$ is less than 8 $\times$ 10$^{9}$ 
M$_{\odot}$. Still smaller masses result for later host types since
M$_{*}$/L$_{\rm K}$ will only decrease for these galaxies. Galaxies of this 
mass are unlikely to host radio-loud AGN as powerful as 
$\approx 10^{24}$ W Hz$^{-1}$, which is the inferred average L$_{\rm 20 cm}$ 
for the non-IRAC detected OIRSs over $0.5 < z < 1.0$.  

On the other hand, if the host galaxies of these OIRSs are $z \approx 1-2$
ellipticals as suggested by Gruppioni et al. (2001), we
would have expected to detect most with IRAC. 
A spinning super-massive black hole is generally held to be a
requirement for a radio-loud active nucleus (Rees 1984). Since the black 
hole mass correlates with the mass of the host galaxy's stellar bulge 
(e.g., Magorrian et al. 1998), radio-loud AGN hosts are known to be 
quite massive, and considerably more luminous than L$^{*}$. 
Using  the above stack-averaged limits, we find the redshift where a massive 
elliptical host galaxy with L$_{\rm K}$ = 3L$_{\rm K}^{*}$ would 
escape detection to be $z \ga 2$, with a corresponding average  L$_{\rm 20 cm}$
of $\approx 10^{25}$ W~Hz$^{-1}$. The OIRSs not detected by IRAC appear to represent
a population of powerful radio galaxies at $z \ga 2$.

This may not be the entire story, of course. The OIRSs lacking IRAC counterparts show a
substantially wider dispersion in $F_{\rm 20 cm}$ compared with the OIRSs detected by IRAC.
Of the four OIRSs with $F_{\rm 20 cm} > 1$ mJy, three lack
IRAC counterparts. At the same time, eight of the thirteen OIRSs fainter than 0.25 mJy
at 20cm  lack IRAC counterparts. It may well be that the non-IRAC detected OIRSs consist
of two distinct populations: one characterized by $L_{\rm 20 cm} < 10^{25}$ W Hz$^{-1}$
and $z \approx 1-2$ (corresponding to the $F_{\rm 20 cm} <$ 0.25 mJy objects) and a much more 
powerful and distant sub-group of radio galaxies (i.e., those with $F_{\rm 20 cm} >$ 1 mJy).
Further progress in understanding the nature of these objects will require
either photometric redshifts derived with deeper {\it Spitzer} IRAC and MIPS 
observations, or through ground-based spectroscopy using the next generation of large 
optical telescopes.

\subsection{Comparison with Sub-Millimeter Galaxies}
It is worth comparing properties of the OIMSs and
OIRSs with other luminous objects at $z \approx 1-4$, in particular, the
galaxies discovered at sub-millimeter wavelengths using instruments like 
SCUBA (Smail et al. 1997). Sub-millimeter galaxies (SMGs) are
extremely luminous (L$_{\rm IR} \approx 10^{13}$ L$_{\odot}$)
and dusty systems, with a redshift distribution that peaks near $z=2.5$ 
(Chapman et al. 2003). Most appear to be powered primarily by star formation.
Indeed, a significant fraction of star formation in the early universe may
take place in them. SMGs also tend to be extremely faint or ``invisible'' at optical 
wavelengths (e.g., Hughes et al. 1998). To ensure meaningful comparisons with 
the OIMSs and OIRSs, we selected a sample of ten SMGs detected at 850 $\mu$m with SCUBA 
in the Lockman Hole East region that were observed with both IRAC 
and MIPS by Egami et al. (2004). Four of the SMGs were also detected at 1.2 mm with
MAMBO (Eales et al. 2003)  by Ivison et al. (2004), and nine were detected at 20 cm
using the VLA's A-configuration (Ivison et al. 2002). The redshift range for the SMG sample 
($z \approx 1-3$) is similar to that of the OIMSs and the IRAC detected OIRSs 
with $z_{\rm phot}$. Rest-frame radio and infrared luminosities as defined in $\S$3.3
were calculated for the SMGs and are listed in Table 5. Likewise, averaged rest-frame
$1-10 \mu$m SEDs were calculated as in $\S$3.2, and are shown in Figure 7 for the ``cold''
SMGs.

Two of the ten SMGs possess power-law SEDs in the IRAC and 
MIPS data, suggesting a significant AGN contribution (``warm'' sources in Egami et al.'s 
terminology). Their power-law exponents ($\alpha = -1.9$) are similar to those of the 
{\it obsSy1} OIMSs. The largest K-band luminosity belongs to the ``warm'' SMG LH850.8b
(L$_{\rm K}$ $= 5.3 \times 10^{12}$ L$_{\odot, \rm K}$), which is 
within a factor of $\approx 2$ of the median L$_{\rm K}$ 
for the hyper-luminous {\it obsSy2} OIMSs ($\approx 10^{13}$ L$_{\odot,\rm K}$).
The remaining eight SMGs have SEDs consistent with
obscured star formation providing the bulk of their luminosity (i.e., ``cold'' sources in
Egami et al.'s terminology). 

There is considerable overlap between the emission properties of the ``cold'' SMGs
and the {\it sb/comb} OIMSs and IRAC detected OIRSs with $z_{\rm phot}$. For all three
groups we derive comparable luminosities at rest-frame K-band (average $L_{\rm K} = 
8 \times 10^{11}$ L$_{\odot,\rm K}$ for the ``cold'' SMGs) and 5 $\mu$m (average 
$\nu$L$_{\nu}$(5 $\mu$m) $\approx 10^{38}$ W), both over an order of 
magnitude smaller than the {\it obsSy1} and {\it obsSy2} OIMSs.
The similarity between the ``cold'' SMGs and the {\it sb/comb} and IRAC detected OIRSs
with $z_{\rm phot}$ extends to their averaged rest-frame $1-10 \mu m$ SEDs as well, as 
Figure 7 shows. In particular, both the ``cold'' SMGs and IRAC detected OIRSs 
with $z_{\rm phot}$ show a prominent 1.6 $\mu$m bump implying substantial evolved 
stellar components. This is followed by a gradual rise to $\lambda_{\rm rest}$$\approx$10 $\mu$m 
as expected for dusty starbursts.  Adopting a starburst M$_{*}$/L$_{\rm K}$ of 0.3, we 
estimate a median stellar mass of M$_{*} = 2 \times 10^{11} 
M_{\odot}$ for the ``cold'' SMGs, which is comparable with the starburst
dominated OIMSs and OIRSs.  Similarly, assuming that the radio continuum
from the ``cold'' SMGs originates in supernova remnants, star
formation rates between 500 and 1700 M$_{\odot}$ yr$^{-1}$ (median SFR =
1100 M$_{\odot}$ yr$^{-1}$) are implied by their L$_{\rm 20 cm}$.
The similar luminosities, implied star formation rates and stellar masses, faint optical 
magnitudes, and SED shapes is evidence that at least a fraction of the OIMSs and IRAC
detected OIRSs are members of the same parent population as the ``cold'' 
(i.e., starburst dominated) SMGs.

The top panel of Figure 8 includes the ten ``warm'' and ``cold'' SMGs in the IRAC 
two-color diagram, along with the OIMSs. As expected, the two ``warm'' SMGs
are positioned in the region occupied by the AGN powered {\it obsSy1}
OIMSs. Neither have mid-IR SEDs that rise as fast as the {\it obsSy2} OIMSs, though
with only two ``warm'' sub-millimeter sources it is impossible to draw any definite 
conclusions. The eight ``cold'' sources are situated close to the dusty M~82 track 
at $z\sim1-3$, both consistent with their redshifts and with obscured star formation 
providing the bulk of their luminosities. There is a clear separation
between the ``cold'' SMGs on one hand and the AGN dominated {\it obsSy1} and {\it obsSy2} 
OIMSs on the other.

Figure 10 shows the relative positions of the OIMSs, OIRSs, and the ten Lockman
Hole East SMGs in a radio/infrared two-color diagram. Systematic trends can be seen:
(1) {\it obsSy1} and
{\it obsSy2} OIMSs extend to the right of center
($F_{\rm 8.0 \mu m}/F_{\rm 3.6 \mu m} \ga 3.5$) due to their
steep power-law SEDs; (2) {\it comb} and {\it sb} OIMSs
are situated primarily to the left of this, with 
$F_{\rm 8.0 \mu m}/F_{\rm 3.6 \mu m}$ ratios between 1.6 
and 4; (3) ``cold'' SMGs are left of center
($F_{\rm 8.0 \mu m}/F_{\rm 3.6 \mu m} < 3$); (4) ``warm'' 
SMGs are to the right of the ``cold'' ones
($F_{\rm 8.0 \mu m}/F_{\rm 3.6 \mu m} > 4$);
(5) the IRAC detected OIRSs primarily occupy the lower-left portion of
the diagram. A few of the latter overlap the region occupied by 
the ``cold'' SMGs. However, the
majority of IRAC detected OIRSs are radio-loud 
(13/19 have {\it q} $< 0$) and are well separated from five of the six
OIRSs with $z_{\rm phot}$. 
It is also apparent that with two exceptions, the OIMSs and 
the ``warm'' and ``cold'' SMGs obey the radio-infrared correlation within
the uncertainties.\footnotemark[14]
\footnotetext[14]{OIMS \#13 is the most radio-loud OIMSs in Houck05, while
the ``cold'' SMG [LE 850] 8a may have strong PAH emission
within the 24 $\mu$m passband, giving rise to its large {\em q}-value.}
OIMSs, OIRSs, and SMGs thus occupy
distinct regions in a radio/infrared two-color diagram, though there is also
considerable overlap, particularly between the OIMSs and SMGs.

This overlap is further illustrated in Figure 11, where L$_{\rm K}$
is plotted against $F_{\rm 8.0 \mu m}/F_{\rm 3.6 \mu m}$. A clear progression 
is seen, with luminous and flat-spectrum ``cold'' SMGs (i.e.,
L$_{\rm K}$ $\approx 10^{11} - 10^{12}$ L$_{\odot,\rm K}$ and $F_{\rm 8.0 \mu m}/F_{\rm 3.6 
\mu m} \sim 1-2$) giving way to {\it obsSy1} 
OIMSs and ``warm'' SMGs (L$_{\rm K}$ $\approx$ few $\times$
$10^{12}$ L$_{\odot,\rm K}$ and $F_{\rm 8.0 \mu m}/F_{\rm 3.6 \mu m} \sim 5-10$),
and finally to the steep power-law {\it obsSy2} OIMSs
(L$_{\rm K}$ $\ga$ 10$^{13}$ L$_{\odot,\rm K}$ and $F_{\rm 8.0 \mu m}/F_{\rm 3.6 
\mu m} > 10$).  This suggests that the OIMSs, a subset of the
IRAC detected OIRSs, and the SMGs are all taken from
the same population of dusty hyper-luminous high-$z$ objects. The chief difference 
between them is the dominant emission mechanism: {\it obsSy1} and {\it obsSy2} OIMSs
in particular can be thought of as analogs of SMGs sources whose luminosities  
in the rest-frame near and mid-IR are dominated by powerful and heavily obscured active
nuclei accreting near the Eddington limit.

Figure 10 also illustrates a dichotomy between the IRAC detected OIRSs with
and without redshifts (i.e., those with and without 24 $\mu$m detections). 
With one exception, all of the former are characterized by
${\it q} > 0$, which is consistent with systems dominated by star formation, a radio-quiet AGN, 
or a combination of the two. By contrast, the remaining IRAC detected OIRSs
all have negative {\it q}-values indicating an increasingly dominant radio-loud
active nucleus. The parent OIRS population is similarly
dominated by radio-loud active nuclei, for excluding the six OIRSs with
$z_{\rm phot}$, 72\% of the sources in Table 1 of Higdon05 have
negative {\it q}. This suggests that the 24 $\mu$m detected OIRSs in Figure 5
are not representative of the larger OIRSs population, but
rather form a starburst dominated sub-group.

\subsection{Do OIMSs and SMGs Form an Evolutionary Sequence?}

Results from the previous sections suggest that the {\it obsSy1}/{\it obsSy2}
OIMSs and ``cold'' SMGs are members of the same parent population of heavily obscured
hyper-luminous galaxies, with the fundamental difference being their dominant
power source, i.e., either accretion near the Eddington limit or star formation.
The {\it obsSy1} and {\it obsSy2} OIMSs in particular can be thought of as heavily 
obscured QSOs. While it is true that very few SMGs contain ``buried''
QSOs ($\approx5\%$, Almaini et al. 2003),  recent ultra-deep X-ray observations 
argue persuasively that many SMGs contain massive 
central black holes and luminous, though not yet dominant, active nuclei. For 
example, $75\%$ of the SMGs with radio counterparts and known redshifts in the 
Chandra Deep Field North were detected in hard X-rays by
Alexander et al. (2005). Their X-ray luminosities (L$_{\rm 0.5-8 ~keV} \approx 10^{36} - 10^{37}$ W), 
levels of obscuration (eighty percent show N$_{\rm H} \approx 10^{20}-10^{24}$ cm$^{-2}$), and 
spectral power-law exponents ($\Gamma \approx 1.8$) are similar to nearby powerful AGN. However,
their X-ray to infrared luminosity ratios (L$_{\rm X}$/L$_{\rm IR} \approx 0.4\%$) are 
typically an order of magnitude smaller than QSOs, consistent with intense
star formation (SFR$\sim1000$ M$_{\odot}$ yr$^{-1}$) providing the bulk of their 
luminosity. This may be simply saying that that the X-ray detected
SMGs are currently accreting at substantially sub-Eddington rates, which would imply that
the only real difference between the {\it obsSy1}/{\it obsSy2} OIMSs and ``cold'' SMGs is the rate at
which material is transfered to the central engine. On the
other hand, near-IR spectra of SMGs typically show that when broad rest-frame optical lines
are present they are typically only $\approx 1000-3000$ km s$^{-1}$ wide (cf. Swinbank et al. 2004),
which is consistent with only {\it modest} black hole masses, i.e., $\la 10^{8}$ M$_{\odot}$. 
It is thus possible (though admittedly speculative) that what distinguishes the AGN powered OIMSs
and ``cold'' SMGs is primarily the central black hole mass rather than simply the accretion rate.

Given the well established correlation between the masses of the spheroid and
central black hole in local galaxies (e.g., Kormendy \& Richstone 1995; Magorrian
et al. 1998) and the evidence linking the SMG population 
with current epoch massive ellipticals (e.g., Smail, Ivison, \& Blain 1997;
Barger et al. 1998; Barger et al. 1999), it is intriguing to consider the
possibility that the {\it obsSy1} and {\it obsSy2} OIMSs represent SMGs that have
made the transition from a starburst dominated ``cold'' phase 
to an accretion dominated ``buried'' QSO phase. A plausible mechanism for this
transition would be the growth of a sufficiently massive black hole (in parallel
with the stellar bulge) during the high-SFR ``cold'' phase, with the 
subsequent transfer of material to the central regions through tidal interactions
or mergers. The energy released during this relatively brief phase would act to 
expell the obscuring gas and dust from the central regions, effectively halting the 
growth of the bulge and super-massive black hole and ``unveiling'' a QSO in the 
manner envisoned by Sanders et al. (1988).

SMGs and OIMSs appear to have have very different mean space densities. If so, this
would have important implications for the envisoned SMG-QSO transition. For example, 
the number of SMGs per square degree with $F_{\rm 850 \mu m} \ge 5$ mJy (which we 
take to represent the hyper-luminous sources) is $\approx$400 using the
cumulative 850 $\mu$m source count formulation in Sanders (2000). 
Assuming they all lie between $z \approx 1-3$, their 
average space density becomes $\rho_{\rm SMG} = 2 \times 10^{-5}$ Mpc$^{-3}$. There are 
seventy sources in the NDWFS Bo\"{o}tes field satisfying $F_{\rm 24 \mu m} > 0.75$ mJy and
$R \ge 24.5$, i.e., OIMSs by our definition, of which fifty-eight have been
observed by {\it Spitzer}. Forty-five of these OIMSs can be confidently classified as 
{\it obsSy1} or {\it obssSy2} on the basis of their IRS spectra and/or steep power-law 
infrared SEDs (Higdon et al. in preparation). Over half of these sources have $z_{\rm spec}
= 0.8-3.1$ on the basis of IRS observations, and the remainder have IRAC/MIPS SEDs 
{\it consistent} with this redshift range.\footnotemark[15] \footnotetext[15]{It was noted 
in $\S$3.2 that the {\it obsSy2} OIMSs show a pronounced
break in their power-law SEDs near $\lambda_{\rm rest} \approx 5-8 \mu$m. For the
{\it obsSy2}-like OIMSs in the larger sample with featureless IRS spectra, the observed 
wavelength of this SED break was found to be consistent with $z\sim1-3$.}
There are thus $\approx50$ {\it obsSy1} and {\it obsSy2} OIMSs in the 8.5 deg$^{2}$ NDWFS.
Assuming that the OIMSs with no redshift measurements are within $z \approx 1-3$,
we estimate a mean space density of $\rho_{OIMS} \approx 
3 \times 10^{-7}$ Mpc$^{-3}$ for this population, or nearly two orders of magnitude
smaller than $\rho_{\rm SMG}$.
This could be simply explained if the duration of the ``buried'' QSO phase is short compared
with the SMG's ``cold'' phase, i.e., the time spent growing a sufficiently massive black hole
and bulge. 

One caveat to the above analysis is the sensitivity of $\rho_{\rm SMG}$
to the adopted 850 $\mu$m flux density cut-off. For example, if only the 
$F_{\rm 850 \mu m} \ge 10$ mJy SMGs pass through a {\it obsSy1} or {\it obsSy2} phase, 
then $\rho_{\rm SMG} \approx 5 \times 10^{-6}$ Mpc$^{-3}$, and $\rho_{\rm SMG}/\rho_{\rm OIMS} 
\approx 17$. The sub-millimeter properties of the OIMS population are largely unknown,
and it is not certain whether OIMSs with $F_{\rm 24 \mu m} > 0.75$ mJy map onto
SMGs with F$_{\rm 850 \mu m} \ge 10$ mJy, or for that matter, SMGs with
F$_{\rm 850 \mu m} \ge 2$ mJy.\footnotemark[16]\footnotetext[16]{For example, none of the 
OIMSs in Houck05 and in the larger sample of OIMSs were detected by MIPS at 70 $\mu$m or 160 $\mu$m. 
Conversely, it had been previously noted that SMGs are fainter than expected at 24 $\mu$m given
their inferred L$_{\rm IR}$ (e.g., Ivison et al. 2004; Egami et al. 2004). This difference 
may simply reflect the presense of a strong contribution from AGN heated dust in the
OIMSs that is lacking in the ``cold'' SMGs.} Despite these uncertainties, it appears likely that 
$\rho_{\rm SMG}/\rho_{\rm OIMS} \approx 20-70$. This suggests that if indeed the {\it obsSy1} and 
{\it obsSy2} OIMSs represent a transition from SMGs to QSOs, and eventually to current epoch
massive elliptical galaxies, it is a relatively rapid one.

\subsection{Do Interactions Trigger the Activity in OIMSs/OIRSs?}

Evidence that the near environment of galaxies plays a role in triggering 
active nuclei has been known for some time.  Heckman et al. (1984, 1986), for
example, found that a large fraction of low redshift quasars have companion
galaxies within $\approx 50 ~kpc$ that typically differ in radial velocity
from the quasar by $\la 1000$ km s$^{-1}$. In addition, roughly 
one-third of radio galaxies show highly peculiar morphologies 
(i.e., tails, bridges, and shells) indicative of tidal interactions. More
recent work supports these results (e.g., Canalizo \& Stockton 2001).
The 3.6 $\mu$m images in Figure 1 suggest that most IRAC detected OIRSs
possess close companions, and may therefore be tidally interacting. If so, this
implies that like low redshift radio galaxies, the central activity in 
the majority of the IRAC detected OIRSs may be induced through collisions.

The fact that most OIMSs appear point-like with IRAC might at first suggest
that they are relatively isolated. However, the mid-IR images shown in 
Figure 3 cannot exclude the possibility that OIMSs possess faint or lower-mass 
companions capable of inducing tidal perturbations large enough to fuel an
active nucleus.  Nor can the IRAC images rule out the possibility that OIMSs 
are a population of late-state mergers. Local examples, such as NGC~7252, typically 
show luminous asymmetric cores with extended low surface brightness tails 
(Schweizer 1982).  Neither would be detectable with IRAC 
given the OIMS's redshifts. It thus remains an interesting possibility that interactions 
have triggered the nuclear activity in both OIRSs and OIMSs. Deep and high 
angular resolution near-infrared imaging studies will be needed to test this
hypothesis.

\section{Conclusions}

OIMSs can be split into two populations based on their rest-frame
1-10 $\mu$m SEDs and IRAC colors in concordance with their IRS spectra: most (12/16) are
dominated by a heavily obscured radio-quiet active nucleus, while the remainder
are powered by either a starburst or a composite starburst/active nucleus. 
The AGN dominated {\it obsSy1} and {\it obsSy2} OIMSs in particular
are extremely luminous in the rest-frame near and mid-IR, with 
$\nu$L$_{\nu}$(5 $\mu$m) comparable to the most luminous local NLRGs and quasars.
They can be regarded as ``buried'' QSOs and likely represent the predecesors
of current epoch massive elliptical galaxies. OIMSs are also distinct from
other high-z source populations routinely selected using UV/optical or 
optical/near-infrared criteria such as LBGs and BzKs in their dominant
power source, levels of obscuration, and mid-IR luminosity. 
Compared with other optically faint mid-IR selected populations (e.g., 
Yan et al. 2007), OIMSs represent extremes in both obscuration and AGN luminosity.
This follows from their respecive selection criteria, which for OIMSs favor steeper
mid-IR continuua and higher levels of extinction. There appears to be
significant overlap between OIMSs and SMGs, with the {\it sb/comb} OIMSs
appearing largely indistinguishable from SMGs in their optical, mid-infrared, and
radio properties. Moreover, the {\it obsSy1} and {\it obsSy2} OIMSs
may represent a brief obscured phase in the transition of a ``cold'' SMG to a
QSO, and eventually to a massive current epoch elliptical galaxy.

The OIRSs do not represent a single source population. The minority (6/35) 
that are detected by both IRAC and MIPS at 24 $\mu$m have SEDs, mid-IR 
colors, and {\it q}-values indicative of either starburst or composite 
starburst/AGN powered systems. For these sources we find $z_{\rm phot}$ in 
between 1.0 and 4.5, implying rest-frame K-band luminosities, maximum star 
formation rates, and stellar masses virtually identical to the ``cold'' (i.e., 
starburst dominated) SMGs in the Lockman Hole East region.

The remaining OIRSs - which comprise $83\%$ of the parent
population in Higdon05 - fall into two classes depending 
upon whether or not they are detected by IRAC. Those that are detected 
have flat mid-IR SEDs implying comparable luminosities from stellar 
photospheres and hot AGN illuminated dust. As a group, their average 
L$_{\rm 24 \mu m}$/L$_{\rm 3.6 \mu m}$ ratio is most consistent with either 
Low Excitation Radio Galaxies (LERGs) or objects
like the ``red and dead'' radio galaxy LBDS 53W091. 
Those that are not detected by IRAC must be at $z \ga 2$ if they
are as massive as the hosts of local radio sources. Both of these groups
are characterized by negative values of {\it q} and thus may
represent a population of {\it relatively} unobscured radio galaxies
at high redshift. Both represent populations that are highly distinct from 
the {\it obsSy1} and {\it obsSy2} OIMSs.

Differences between the optically {\it invisible} populations detected through 
observations at sub-millimeter, mid-IR and radio wavelengths can be understood 
in terms of selection effects: (1) extreme optical/mid-IR luminosity ratios appears 
to ensure highly obscured AGN dominated sources ({\it obsSy1} and {\it obsSy2} OIMSs), 
(2) extreme optical/sub-millimeter luminosity ratios will select highly obscured 
sources primarily powered by star formation (SCUBA/MAMBO sources), and 
(3) compact sub-mJy radio sources lacking optical
counterparts appears to preferentially choose distant BL Lac-like objects (OIRSs).

We find no evidence that OIMSs or OIRSs inhabit the cores of rich clusters.
Nor do we find significant differences in local galaxy density between the
two on $\la 100$ kpc scales. However, unlike the OIMSs, a large fraction of the 
IRAC detected OIRSs appear to possess close and massive companions,
though higher angular resolution studies will be needed to 
reach firm conclusions. This suggests that the luminosity of OIRSs (and conceivably
OIMSs) may be triggered by tidal interactions, as appears to be the case for 
low-$z$ radio galaxies and quasars. The IRAC detected OIRSs may thus represent the formation
of very massive galaxies at high redshift through major mergers.

\acknowledgements

We wish to thank Terry Herter for valuable discussions relating to the
template fitting routines, as well as Julien Devriendt and Kevin Xu for 
access to their libraries of galaxy SEDs. We also wish to thank the anonymous
referee for helpful suggestions and comments. This work is
based in part on observations made with the {\em Spitzer Space Telescope},
which is operated by the Jet Propulsion Laboratory, California
Institute of Technology under a contract with NASA. Support for this
work was provided by NASA through awards issued by JPL/Caltech.
This research has made use
of the NASA/IPAC Extragalactic Database (NED) which is operated by
the Jet Propulsion Laboratory, California Institute of Technology,
under contract with the National Aeronautics and Space
Administration.

\newpage
\references

\reference{} Adelberger, K., Steidel, C., Shapley, A., Hunt, M., Erb, D., Reddy, N., 
	     \& Pettini, M.  2004, \apj, 607, 226

\reference{} Alexander, D., Bauer, F., Chapman, S., Smail, I., Blain, A., Brandt, W..
	     \& Ivison, R. 2005, \apj, 632, 736

\reference{} Allen, S., Dunn, R., Fabian, A., Taylor, G. \& Reynolds, C. 2006, \mnras, 372, 21

\reference{} Almaini, O. et al. 2003, \mnras, 338, 303

\reference{}  Appleton, P. et al.  2004, \apjs, 154, 147

\reference{} Barger, A., Cowie, L., Sanders, D., Fulton, E., Taniguchi, Y., Sato, Y.,
	   Kawara, K., \& Okuda, H. 1998, \nat, 394, 248

\reference{} Barger, A., Cowie, L., \& Sanders, B. 1999, \apj, 518, L5

\reference{} Barr, J., M., Baker, J. C., Bremer, M. N., Hunstead, R. W., \&
	   Bland-Hawthron, J. 2004, \aj, 128, 2660

\reference{} Bell, E. F. \& de Jong, R. S. 2001, \apj, 550,  212

\reference{} Bell, E. F.  2003, \apj, 586, 794

\reference{} Bertin, E.~\& Arnouts, S.\ 1996, \aaps, 117, 393 

\reference{} Bessel, M. S. 1979, \pasp, 91, 589

\reference{} Best, P. N., Lehnert, M. D., Miley, G. K., \& R\"{o}ttgering, H. J. 2003, \mnras, 343, 1

\reference{} Canalizo, G. \& Stockton, A. 2001, \apj, 555, 719

\reference{} Chapman, S., Blain, A., Ivison, R. \& Smail, I. 2003, \nat, 422, 695

\reference{} Chary, R., \& Elbaz, D.  2001, \apj, 556, 562

\reference{} Daddi, E., Cimatti, A., Renzini, A., Fontana, A., Mignoli, M., Pozzetti, L.
                     Tozzi, P. \& Zamorani, G. 2004, \apj, 617, 746

\reference{} Devriendt, J. E. G., Guiderdoni, B., \& Sadat, R. 1999,  \aap, 350, 381

\reference{} de Vries, W. H., Morganti, R., R{\"o}ttgering, H. J., Vermeulen, R.,
                     van Breugel, W., Rengelink, R., \& Jarvis, M. J.  2002, \aj, 123, 1784

\reference{} Eales, S. A., Bertoldi, F., Ivison, R. J., Carilli, C., Dunne, L,
	   \& Owen, F. 2003,\mnras, 344, 169

\reference{} Eckart, A., van der Werf, P., Hofmann, R., \& Harris, A. 1994, \apj, 424, 627

\reference{} Egami, E. et al. 2004, \apjs, 154, 130

\reference{} Eisenhardt, P. R. et al. 2004, \apjs, 154, 48

\reference{} Elvis, M., Wilkes, B., McDowell, J., Green, R., Bechtold, J., Willner, S.,
                     Oey, M., Polomski, E. \& Cutri, R. 1994, \apjs, 95, 1 

\reference{} Evans, D., Worrall, D., Hardcastle, M., Kraft, R., \& Birkinshaw, M. 2006,
	   \apj, 642, 96

\reference{} Fazio, G. G. et al. 2004, \apjs, 154, 10

\reference{} Fomalont, E., Kellermann, K., Partridge, R., \& Richards, E. 2002, \aj, 123, 2402

\reference{} Gruppioni, C., Oliver, S., \& Serjeant, S.\ 2001, \apss, 276, 791 

\reference{} Hardcastle, M., Evans, D., \& Croston, J. 2006, \mnras, 370, 1893

\reference{} Heckman, T. M., Bothun, G. D., Balick, B., \& Smith, E. P. 1984, \aj, 89, 958

\reference{} Heckman, T. M. et al. 1986, \apj, 311, 526

\reference{} Higdon, J. L., Higdon, S. J. U., Weedman, D. W., Houck, J. R.,
                     Le Floc'h, E., Brown, M. J. I., Dey, A., Jannuzi, B. T., Soifer, B. T.
                     \& Rieke, M. J.  2005, \apj, 626, 58 (Higdon05)

\reference{} Houck, J. R. et al. 2004, \apjs, 154, 18

\reference{} Houck, J. R., Soifer, B. T., Weedman, D., Higdon, S. J. U.,
                    Higdon, J. L., Herter, T., Brown, M. J. I., Dey, A.,
                    Jannuzi, B. T., Le Floc'h, E., Rieke, M., Armus, L.,
                   Charmandaris, V., \& Teplitz, H.  2005, \apj, 622, 105 (Houck05)

\reference{} Hu, E. M., \& Ridgeway, S. E. 1994, \aj, 107, 1303

\reference{} Hughes, D.~H., et al.\ 1998, \nat, 394, 241 

\reference{} Ivison, R. J. et al. 2002, \mnras, 337, 1

\reference{} Ivison, R. J. et al. 2004, \apjs, 154, 124

\reference{} Jannuzi, B. \& Dey, A.  1999, ASP Conference Series Vol.~191, 
                     (R. Weymann, L. Storrie-Lombardi, M. Sawicki, \&
                     R. Brunner, eds.), p. 111

\reference{} Kenter, A., Murray, S., Forman, W., Jones, C., Green, P., Kochanek, C.,
	  Vikhlinin, A., Fabricant, D., Fazio, G., Brand, K., Brown, M., Day, A.,
	  Jannuzi, B., Najita, J., McNamara, B., Shields, J., \& Rieke, M. 2005,
	  \apjs, 161, 9

\reference{} Kharb, P., Shastri, P., \& Gabuzda, D. 2005, \apj, 632, 69

\reference{} Kochanek, C. S., Pahre, M., Falco, E., Huchra, J.,
                    Mader, J., Jarrett, T., Chester, T., Cutri, R., \& Schneider, S.
                    2001, \apj, 560, 566

\reference{} Kormendy, J. \& Richstone, D.  1995, \araa, 33, 581

\reference{} Lacy, M. et al. 2004, \apjs, 154, 166

\reference{} Laing, R., Riley, J., \& Longair, M. 1983, \mnras, 204, 151

\reference{} Laing, R., Jenkins, C., Wall, J., \& Unger, S. 1994, ASP Conference Series, 
	   Vol.~54, (Bicknell, Dopita, and Quinn, eds.), p. 56

\reference{} Lester, D., Carr, J., Joy, M., \& Gaffney, N.  1990, \apj, 352, 544

\reference{} Magorrian, J. et al. 1998, \aj, 115, 2285

\reference{} Mathis, J. C. 1990, \araa, 28, 37

\reference{} McCarthy, P. 1993, \araa, 31, 639

\reference{} Ogle, P., Whysong, D. \& Antonucci, R. 2006, \apj, 647, 161

\reference{} Persic, M. et al. 2004, \aap, 419, 849

\reference{} Rees, M. J. 1984, \araa, 22, 471

\reference{} Richards, E., Fomalont, E., Kellermann, K., Windhorst, R., Partridge, R.,
	   Cowie, L. \& Barger, A. 1999, \apj, 526, L73

\reference{} Rieke, G. H. et al. 2004, \apjs, 154, 25

\reference{} Sanders, D., Soifer, B., Elias, J., Madore, B., Matthews, K., Neugebauer, G. \&
	   Scoville, N. 1988, \apj, 325, 74

\reference{} Sanders, D. 2000, Advances in Space Research, 25, 2251

\reference{} Schweizer, F. 1982, \apj, 252, 455

\reference{} Siebenmorgen, R., Haas, M., Kr\"{u}gel, E., \& Schulz, B. 2005, \aap, 436, L5

\reference{} Silva, L., Granato, G., Bressan, A., \& Danese, L. 1998, \apj, 509, 103

\reference{} Smail, I., Ivison, R. J., \& Blain, A. W. 1997, \apj, 490, 5

\reference{} Soifer, B. T. et al. 2004, \apjs, 154, 151

\reference{} Stern, D., Chary, R., Eisenhardt, P., \& Moustakas, L. 2006, \apj, 132, 1405

\reference{} Steidel, C., Giavalisco, M., Dickinson, M. \& Adelberger, K. 1996, \aj, 112, 352

\reference{} Swinbank, A., Smail, I., Chapman, S., Blain, A., Ivison, R. \& Keel, R.  2004, \apj, 617, 64

\reference{} Telesco, C. M., Campinis, H., Joy, M., Dietz, K., \& Decher, R.  1991, \apj, 369, 135

\reference{} Tokunaga, A. T., Simons, D. A., \& Vacca, W. D. 2002, \pasp, 114, 180

\reference{} Werner, M. W. et al. 2004, \apjs, 154, 1

\reference{}  Xu, C., Lonsdale, C.~J., Shupe, D.~L., O'Linger, J., 
                      \& Masci, F.  2001, \apj, 562, 179 

\reference{} Yan, L., Sajina, A., Fadda, D., Choi, P., Armus, L., Helou, G., Teplotz, H.
	   Frarer, D., \& Surace, J.  2007, \apj, 658, 778

\newpage
\clearpage
\begin{deluxetable}{ccccccccccc}
\tablenum{1}
\tablecaption{Observed Flux Densities For The IRAC Detected OIRSs}
\tablewidth{0pc}
\tabletypesize{\scriptsize}
\tablehead{
\colhead{OIRS\tablenotemark{(a)}} & \colhead{F$_{B_W}$\tablenotemark{(b)}} & \colhead{F$_{R}$\tablenotemark{(b)}} & \colhead{F$_{I}$\tablenotemark{(b)}} &
\colhead{F$_{\rm 3.6 \mu m}$ } & \colhead{F$_{\rm 4.5 \mu m}$} & \colhead{F$_{\rm 5.8 \mu m}$} & \colhead{F$_{\rm 8.0 \mu m}$} &
\colhead{F$_{\rm 24 \mu m}$} &  \colhead{F$_{\rm 20 cm}$}  & \colhead{ q\tablenotemark{(c)} }\\
\colhead{[HHW2005]~$\#$} & \colhead{ ($\mu$Jy) } &  \colhead{ ($\mu$Jy) } & \colhead{ ($\mu$Jy) } &
\colhead{($\mu$Jy) } & \colhead{($\mu$Jy) } &\colhead{($\mu$Jy) } & \colhead{($\mu$Jy) } &
\colhead{ (mJy) } &  \colhead{ (mJy) } &  }
\startdata
 ~19 & $<$0.062 & $<$0.256 & $<$0.404 & 26.7 $\pm$ 1.9  & 27.6 $\pm$ 2.5  &  25.0 $\pm$ 12.5 &   $<$35.0        & $<$0.18          & 0.57 $\pm$ 0.04 & $<$-0.50 \\
 ~49 & $<$0.056 & $<$0.233 & $<$0.255 & 12.5 $\pm$ 2.1  & 15.6 $\pm$ 2.7  &   $<$40.5        &  33.2 $\pm$ 12.1 & $<$0.18          & 0.57 $\pm$ 0.03 & $<$-0.50\\
 ~79 & $<$0.081 & $<$0.256 & $<$0.193 & 29.1 $\pm$ 2.1  & 35.2 $\pm$ 2.7  &   $<$45.3        &  43.9 $\pm$ 12.2 & $<$0.18          & 0.95 $\pm$ 0.04 & $<$-0.72\\
 ~97 & $<$0.056 & $<$0.147 & $<$0.212 & 19.7 $\pm$ 1.9  & 24.1 $\pm$ 2.5  &   $<$40.7        &  30.2 $\pm$ 12.1 & 0.23 $\pm$ 0.04  & 0.40 $\pm$ 0.03 & -0.24 $\pm$ 0.08\\
 114 & $<$0.081 & $<$0.194 & $<$0.255 & 16.8 $\pm$ 1.8  & 18.4 $\pm$ 2.3  &  31.3 $\pm$ 12.6 &  31.6 $\pm$ 11.2 & $<$0.18          & 0.37 $\pm$ 0.05 & $<$-0.31\\
 156 & $<$0.067 & $<$0.122 & $<$0.176 & 22.0 $\pm$ 1.9  & 22.8 $\pm$ 2.5  &   $<$39.9        &   $<$35.8        & $<$0.18          & 0.80 $\pm$ 0.06 & $<$-0.65\\
 176 & $<$0.067 & $<$0.194 & $<$0.336 & 38.2 $\pm$ 2.2  & 74.2 $\pm$ 3.1  & 155.3 $\pm$ 15.1 & 248.1 $\pm$ 13.2 & 0.46 $\pm$ 0.04  & 0.25 $\pm$ 0.04 & 0.26 $\pm$ 0.08\\
 182 & $<$0.067 & $<$0.161 & $<$0.336 &  9.7 $\pm$ 1.7  &  8.7 $\pm$ 1.8  &   $<$39.9        &   $<$36.0        & $<$0.18          & 0.55 $\pm$ 0.04 & $<$-0.49\\
 185 & $<$0.067 & $<$0.122 & $<$0.212 & 16.0 $\pm$ 1.9  & 17.0 $\pm$ 2.4  &  31.0 $\pm$ 13.0 &   $<$35.8        & $<$0.18          & 0.33 $\pm$ 0.04 & $<$-0.26\\
 208 & $<$0.067 & $<$0.122 & $<$0.193 & 32.1 $\pm$ 2.1  & 35.8 $\pm$ 1.7  &   $<$40.3        &  39.8 $\pm$ 12.1 & 0.26 $\pm$ 0.08  & 0.19 $\pm$ 0.03 & 0.13 $\pm$ 0.15\\
 232 & $<$0.074 & $<$0.281 & $<$0.336 &  8.1 $\pm$ 1.7  &  9.1 $\pm$ 2.3  &   $<$40.4        &   $<$35.9        & $<$0.18          & 0.51 $\pm$ 0.02 & $<$-0.45\\
 245 & $<$0.074 & $<$0.256 & $<$0.306 & 21.2 $\pm$ 1.9  & 25.7 $\pm$ 2.5  &  41.5 $\pm$ 12.8 &  25.0 $\pm$ 11.2 & 0.44 $\pm$ 0.05  & 0.24 $\pm$ 0.03 & 0.26 $\pm$ 0.07\\
 278 & $<$0.076 & $<$0.148 & $<$0.146 &     $<$2.2      &  6.1 $\pm$ 1.6  &  29.4 $\pm$ 13.5 &   $<$35.6        & $<$0.18          & 0.45 $\pm$ 0.07 & $<$-0.40\\ 
 346 & $<$0.067 & $<$0.112 & $<$0.176 &  8.8 $\pm$ 1.6  & 13.2 $\pm$ 2.2  &   $<$36.8        &  47.9 $\pm$ 11.4 & $<$0.18          & 0.37 $\pm$ 0.04 & $<$-0.31\\
 349 & $<$0.062 & $<$0.134 & $<$0.232 & 26.7 $\pm$ 0.9  & 28.9 $\pm$ 1.5  &  22.5 $\pm$ 6.9  &  38.3 $\pm$ 7.2  & 0.25 $\pm$ 0.08  & 0.24 $\pm$ 0.05 & 0.02 $\pm$ 0.17\\
 363 & $<$0.056 & $<$0.134 & $<$0.193 &  8.6 $\pm$ 0.9  & 12.0 $\pm$ 1.4  &   $<$40.3        &   $<$36.0        & 0.38 $\pm$ 0.05  & 0.15 $\pm$ 0.04 & 0.40 $\pm$ 0.13\\
 380 & $<$0.074 & $<$0.194 & $<$0.279 &  9.4 $\pm$ 1.7  & 13.5 $\pm$ 2.4  &   $<$40.0        &   $<$35.8        & $<$0.18          & 1.92 $\pm$ 0.04 & $<$-1.0\\
 389 & $<$0.074 & $<$0.112 & $<$0.193 & 12.4 $\pm$ 1.8  & 22.3 $\pm$ 2.5  &   $<$40.0        &   $<$36.0        & $<$0.18          & 0.30 $\pm$ 0.02 & $<$-0.22\\
 410 & $<$0.067 & $<$0.147 & $<$0.193 & 28.2 $\pm$ 2.0  & 32.5 $\pm$ 2.6  &   $<$40.3        &   $<$36.0        & $<$0.18          & 0.88 $\pm$ 0.08 & $<$-0.69
\enddata

\tablenotetext{(a)}{OIRS number from Table 2 in Higdon05, where source positions
                    are given. These sources are also designated [HHW2005] by SIMBAD as (e.g.,
                    [HHW2005]~97).}
\tablenotetext{(b)}{ $3 \sigma$ limits from NDWFS images.}
\tablenotetext{(c)}{ $q \equiv \log(F_{24~\micron}/F_{20~\rm cm}$).
The radio flux densities are from Higdon05, while the 24~$\mu$m flux
densities and limits are from this work.}

\end{deluxetable}

\begin{deluxetable}{ccccccccccc}
\tablenum{2}
\tablecaption{Observed Flux Densities For The OIMSs}
\tablewidth{0pc}
\tabletypesize{\scriptsize}
\tablehead{
\colhead{OIMS\tablenotemark{(a)}}  & \colhead{F$_{B_W}$\tablenotemark{(b)}} & 
\colhead{F$_{R}$\tablenotemark{(b)}} & \colhead{F$_{I}$\tablenotemark{(b)}} &
\colhead{F$_{\rm 3.6 \mu m}$} & \colhead{F$_{\rm 4.5 \mu m}$} & \colhead{F$_{\rm 5.8 \mu m}$} & \colhead{F$_{\rm 8.0 \mu m}$} & 
\colhead{F$_{\rm 24 \mu m}$}  & \colhead{F$_{\rm 20 cm}$\tablenotemark{(c)}}  & \colhead{ q\tablenotemark{(d)} }\\
\colhead{[HSW2005]~$\#$} & \colhead{($\mu$Jy)} &\colhead{($\mu$Jy)} &\colhead{($\mu$Jy)} &\colhead{($\mu$Jy)} &\colhead{($\mu$Jy)} &
\colhead{($\mu$Jy)} &\colhead{($\mu$Jy)} &\colhead{(mJy)} &\colhead{(mJy)} & \colhead{ } } 
\startdata
~1 &  ~~0.169 &  ~~0.337 &  ~~0.443 & 26.9 $\pm$ 1.0 & ~46.0 $\pm$ 1.4 & ~79.3 $\pm$ ~8.9 & 207.7 $\pm$ ~8.2 & 1.24 $\pm$ 0.01 &$<$0.15 & $>$0.90\\
~2 & $<$0.081 & $<$0.102 &  ~~0.404 & 17.0 $\pm$ 1.0 & ~26.0 $\pm$ 1.4 & ~20.4 $\pm$ ~8.0 & ~62.1 $\pm$ ~7.3 & 0.89 $\pm$ 0.01 &$<$0.15 & $>$0.76\\
~3 & $<$0.067 & $<$0.123 & $<$0.193 & 22.3 $\pm$ 1.0 & ~26.0 $\pm$ 1.6 & ~39.0 $\pm$ ~7.4 & ~73.9 $\pm$ ~8.2 & 1.81 $\pm$ 0.02 &~~0.31 $\pm$ 0.03  &~~0.77 $\pm$ 0.04\\
~4 &  ~~0.089 &  ~~0.445 &  ~~0.584 & 17.6 $\pm$ 0.9 & ~38.0 $\pm$ 1.4 & ~92.4 $\pm$ ~6.4 & 248.5 $\pm$ ~7.3 & 1.08 $\pm$ 0.01 &~~0.24 $\pm$ 0.04  &~~0.65 $\pm$ 0.06\\
~5 & $<$0.067 & $<$0.123 & $<$0.193 & ~4.8 $\pm$ 0.8 & ~~9.4 $\pm$ 1.3 &        $<$33.0   &    $<$21.0       & 0.87 $\pm$ 0.01 &$<$0.15 & $>$0.75\\
~6 & $<$0.067 & $<$0.123 & $<$0.193 & 15.6 $\pm$ 0.8 & ~21.1 $\pm$ 1.5 & ~30.4 $\pm$ ~6.3 & ~92.6 $\pm$ ~8.2 & 1.03 $\pm$ 0.02 &$<$0.15 & $>$0.82\\
~7 &  ~~0.141 & $<$0.123 &  ~~0.443 & 41.7 $\pm$ 0.8 & ~55.3 $\pm$ 1.4 & ~70.1 $\pm$ ~6.5 & 124.7 $\pm$ ~7.8 & 0.78 $\pm$ 0.01 &$<$0.15 & $>$0.70\\
~8 &  ~~0.129 &  ~~0.123 &  ~~0.279 & 12.2 $\pm$ 0.8 & ~18.8 $\pm$ 1.4 & ~69.6 $\pm$ ~6.7 & 231.3 $\pm$ ~7.9 & 2.65 $\pm$ 0.02 &$<$0.15 & $>$1.23\\
~9 & $<$0.067 & $<$0.123 & $<$0.193 & ~7.4 $\pm$ 1.0 & ~28.3 $\pm$ 1.2 & ~92.4 $\pm$ ~6.8 & 533.3 $\pm$ ~5.9 & 3.83 $\pm$ 0.02 &~~0.42 $\pm$ 0.03 &~~0.96 $\pm$ 0.04\\
10 &  ~~0.511 &  ~~0.123 &  ~~0.336 & ~6.0 $\pm$ 0.8 & ~12.5 $\pm$ 1.4 & ~25.3 $\pm$ ~7.7 &    $<$21.0       & 1.45 $\pm$ 0.01 &$<$0.15 & $>$0.97\\
12 & $<$0.067 & $<$0.123 & $<$0.193 & 14.3 $\pm$ 1.0 & ~25.9 $\pm$ 1.3 &        $<$41.0   & ~38.0 $\pm$ ~6.8 & 1.12 $\pm$ 0.01 &~~0.20 $\pm$ 0.03 &~~0.80 $\pm$ 0.07\\
13 &  ~~0.561 &  ~~0.929 &  ~~1.112 & 42.1 $\pm$ 3.2 & 107.2 $\pm$ 4.5 & 328.1 $\pm$ 22.5 & 699.2 $\pm$ 25.1 & 2.30 $\pm$ 0.01 &~~5.09 $\pm$ 0.03 &~-0.34 $\pm$ 0.01\\
14 & $<$0.067 & $<$0.123 & $<$0.193 & 16.9 $\pm$ 0.8 & ~22.6 $\pm$ 1.5 & ~26.4 $\pm$ ~7.0 & ~25.7 $\pm$ ~7.8 & 0.79 $\pm$ 0.01 &$<$0.15 & $>$0.69\\
15 & $<$0.067 & $<$0.123 & $<$0.193 & 33.3 $\pm$ 0.9 & ~48.5 $\pm$ 1.4 & ~83.6 $\pm$ ~6.9 & 148.8 $\pm$ ~7.5 & 1.05 $\pm$ 0.01 &$<$0.15 & $>$0.82\\
16 & $<$0.067 & $<$0.123 & $<$0.193 & ~3.6 $\pm$ 0.9 & ~~9.0 $\pm$ 1.4 & ~16.0 $\pm$ ~7.5 & ~60.4 $\pm$ ~8.0 & 1.04 $\pm$ 0.01 &$<$0.15 & $>$0.81\\
17 & $<$0.067 & $<$0.123 & $<$0.193 & 20.6 $\pm$ 0.8 & \nodata\tablenotemark{(e)}& ~56.2 $\pm$ ~6.3 &\nodata\tablenotemark{(e)} & 1.23 $\pm$ 0.01 &$<$0.15 & $>$0.89
\enddata
\tablenotetext{(a)}{ OIMS number from Table~1 in Houck05, where source positions are given.
                    These sources are also designated [HSW2005] by SIMBAD (e.g., 
                    [HSW2005] 13).}
\tablenotetext{(b)}{ Optical flux densities or $3 \sigma$ limits from NDWFS.}
\tablenotetext{(c)}{ 20 cm flux densities or $5 \sigma$ limits from de Vries et al. (2002).}
\tablenotetext{(d)}{{\em q} $\equiv$ log($F_{\rm 24 \mu m}/F_{\rm 20cm}$). The $F_{\rm 24 \mu m}$
values are from Houck05.}
\tablenotetext{(e)}{ No data at these bands.}
\end{deluxetable}


\newpage
\clearpage
\begin{deluxetable}{cccccccc}
\tablenum{3}
\tablecaption{Derived OIRS $z_{\rm phot}$ And Luminosities}
\tablewidth{0pc}
\tabletypesize{\scriptsize}
\tablehead{  
\colhead{ OIRS\tablenotemark{(a)} } &  
\colhead{$z_{\rm phot}$} &  
\colhead{Template\tablenotemark{(b)}}  &  
\colhead{A$_{\rm V}$\tablenotemark{(c)}} &  
\colhead{L$_{\rm K}$\tablenotemark{(d)}} &  
 \colhead{L$_{\rm 5 \mu m}$\tablenotemark{(e)}} &
\colhead{M$_{*}$\tablenotemark{(f)}} &   
\colhead{L$_{\rm 20 cm}$}   \\
\colhead{[HHW2005]~$\#$} &  
\colhead{  } &  
\colhead{  }  &
\colhead{  } &  
\colhead{ ( 10$^{11}$ L$_{\odot,\rm K}$ )  } & 
\colhead{ ( 10$^{37}$ W ) } &  
\colhead{ ( 10$^{11}$ M$_{\odot}$ ) } &  
\colhead{ ( 10$^{25}$ W Hz$^{-1}$ ) }
}
\startdata
 ~97 & 1.5 $\pm$ 0.3 &  Arp~220  & 2 & ~~~3.7 $\pm$ 1.6  &  ~0.3 $\pm$ 0.2  & ~~~1.4 $\pm$ 0.5  & 0.4 $\pm$ 0.2\\
 176 & 4.5 $\pm$ 0.2 &   M~82    & 1 & ~~278. $\pm$ 26.  &  11.6 $\pm$ 1.3   & ~~116. $\pm$ 11. & 2.9 $\pm$ 0.3\\
 208 & 2.0 $\pm$ 0.3 &   M~82    & 1 & ~~~9.9 $\pm$ 3.4  &  ~0.7 $\pm$ 0.2  & ~~~3.4 $\pm$ 1.1  & 0.4 $\pm$ 0.1\\
 245 & 2.0 $\pm$ 0.3 &  Arp~220  & 2 & ~~~7.4 $\pm$ 2.7  &  ~1.2 $\pm$ 0.3  & ~~~3.3 $\pm$ 1.1  & 0.5 $\pm$ 0.2\\
 349 & 1.0 $\pm$ 0.2 &   M~82    & 2 & ~~~1.8 $\pm$ 0.6  &  0.09 $\pm$ 0.03 & ~~~0.8 $\pm$ 0.3  & 0.2 $\pm$ 0.1\\
 363 & 2.2 $\pm$ 0.6 &  Arp~220  & 1 & ~~~4.5 $\pm$ 2.4  &  ~1.1 $\pm$ 0.6  & ~~~2.3 $\pm$ 1.4  & 0.4 $\pm$ 0.2\\
\enddata
\tablenotetext{(a)}{OIRS number from Table 2 in Higdon05, where source positions
                    are given. These objects are designated in SIMBAD as [HHW2005] $\#$ (e.g.,
                    [HHW2005]~97).}
\tablenotetext{(b)}{The best fitting galaxy template; see $\S$3.1 for descriptions.}
\tablenotetext{(c)}{Additional visual extinction in magnitudes to satisfy optical constraints.}
\tablenotetext{(d)}{The source rest-frame Johnson K-band luminosity, defined as 
                    L$_{\rm K}$ = 4$\pi$D$_{\rm L}^2$(1+z)$^{-1}F_{\nu} \Delta\nu_{\rm K}$,
                    in units of the sun's K-band luminosity, L$_{\odot, \rm K}$ = 8.7 $\times$ 10$^{24}$ W  
                    (Bessel 1979). D$_{\rm L}$ is the luminosity distance, $\Delta\nu_{\rm K}$ is the
                    Johnson K-band filter FWHM ($2.1 \times 10^{13}$ Hz; Tokunaga et al. 2002), and
                    $F_{\nu}$ is the source flux density at $\lambda_{\rm obs} = (1 + z) 2.2 \mu m$
                    estimated from the source's SED template.}
\tablenotetext{(e)}{The source rest-frame monochromatic luminosity at 5 $\mu$m in Watts, defined as $\nu$L$_{\nu}$(5 $\mu$m),
                    where L$_{\nu}$(5 $\mu$m) = 4$\pi$D$_{\rm L}^2$(1+z)$^{-1}F_{\nu}$.
                    D$_{\rm L}$ is again the luminosity distance and $F_{\nu}$ is the source flux density
	 	    at $\lambda_{\rm obs} = (1 + z) 5 \mu m$ estimated from the source's SED template.}
\tablenotetext{(f)}{Stellar mass derived by multiplying L$_{\rm K}$ by M$_{*}$/L$_{\rm K}$ =
                    0.3, which is appropriate for a starburst galaxy (Bell \& de Jong 2001). }
\end{deluxetable}

\begin{deluxetable}{ccccccccc}
\tablenum{4}
\tablecaption{Derived OIMS Luminosities}
\tablewidth{0pc}
\tabletypesize{\scriptsize}
\tablehead{  
\colhead{OIMS\tablenotemark{(a)} } &  
\colhead{$z_{\rm spec}$} &  
\colhead{Template\tablenotemark{(b)}}  &    
\colhead{L$_{\rm K}$\tablenotemark{(c)}} &   
\colhead{L$_{\rm 5 \mu m}$\tablenotemark{(d)}} &
\colhead{M$_{*}$\tablenotemark{(e)}} &  
\colhead{L$_{\rm 20 cm}$}  & 
\colhead{log L$_{\rm IR}$\tablenotemark{(f)} }  \\
\colhead{[HSW2005]~$\#$} &  
\colhead{  } &  
\colhead{  }  &
\colhead{ ( 10$^{12}$ L$_{\odot, \rm K}$ )  } & 
\colhead{ ( 10$^{38}$ W ) } & 
\colhead{ ( 10$^{11}$ M$_{\odot}$ ) }  &  
\colhead{ ( 10$^{25}$ W Hz$^{-1}$ ) }  & 
\colhead{ ( L$_{\odot}$ ) }
}
\startdata
~1 & 2.64 $\pm$ 0.25 &obsSy2&   ~8.3 $\pm$ 1.8 & ~8.3 $\pm$ 1.8  & $<$25.           & $<$0.8           &  13.3  \\
~2 & 1.86 $\pm$ 0.07 &sb    &   ~0.7 $\pm$ 0.1 & ~1.0 $\pm$ 0.1  & ~~1.9 $\pm$ 0.2  & $<$0.4           &  13.0  \\
~3 & 1.78 $\pm$ 0.30 &obsSy1&   ~0.9 $\pm$ 0.3 & ~0.9 $\pm$ 0.3  & ~$<$2.7          &  ~~0.5 $\pm$ 0.2 &  13.0  \\
~4 & 2.59 $\pm$ 0.34 &obsSy2&   ~9.4 $\pm$ 2.9 & ~9.4 $\pm$ 2.4  & $<$28.           &  ~~0.9 $\pm$ 0.3 &  13.2  \\
~5 & 2.34 $\pm$ 0.28 &obsSy2& \nodata          & \nodata         & \nodata          & $<$0.7           &  13.0  \\
~6 & 1.96 $\pm$ 0.34 &obsSy1&   ~1.2 $\pm$ 0.5 & ~1.7 $\pm$ 0.6  & ~$<$3.7          & $<$0.4           &  12.8  \\
~7 & 1.78 $\pm$ 0.14 &comb  &   ~1.6 $\pm$ 0.3 & ~1.4 $\pm$ 0.3  & ~~4.8 $\pm$ 0.9  & $<$0.3           &  13.4  \\
~8 & 2.62 $\pm$ 0.26 &obsSy2&   ~9.1 $\pm$ 2.1 & 10.2 $\pm$ 2.2  & $<$27.           & $<$0.8           &  13.6  \\
~9 & 2.46 $\pm$ 0.20 &obsSy2&   16.3 $\pm$ 3.2 & 21.0 $\pm$ 4.1  & $<$49.           &  ~~1.4 $\pm$ 0.3 &  13.7  \\
10 & 2.08 $\pm$ 0.21 &obsSy1&   ~0.9 $\pm$ 0.2 & ~0.8 $\pm$ 0.2  & ~$<$2.8          & $<$0.5           &  13.0  \\
12 & 2.13 $\pm$ 0.09 &comb  &   ~0.7 $\pm$ 0.1 & ~0.4 $\pm$ 0.1  & ~~2.1 $\pm$ 0.2  &  ~~0.5 $\pm$ 0.1 &  13.8  \\
13 & 1.95 $\pm$ 0.17 &obsSy2&   10.2 $\pm$ 2.1 & 10.9 $\pm$ 2.2  & $<$31.           &  ~10.2 $\pm$ 2.1 &  13.3  \\
14 & 2.26 $\pm$ 0.11 &comb  &   ~0.8 $\pm$ 0.8 & ~3.0 $\pm$ 0.4  & ~~2.4 $\pm$ 0.3  & $<$0.6           &  13.7  \\
15 & 1.75 $\pm$ 0.21 &obsSy1&   ~1.8 $\pm$ 0.5 & ~1.6 $\pm$ 0.4  & ~$<$5.           & $<$0.3           &  12.7  \\
16 & 2.73 $\pm$ 0.19 &obsSy2&   ~2.7 $\pm$ 0.4 & ~2.9 $\pm$ 0.5  & $<$8.            & $<$0.8           &  13.2  \\
17 & 2.13 $\pm$ 0.17 &obsSy1& \nodata          & \nodata         & \nodata          & $<$0.5           &  13.0 
\enddata
\tablenotetext{(a)}{Source number from Table~1 of Houck05 where coordinates can be found.
                    The OIMSs can also be referenced by their SIMBAD designation HSW2005 $\#$
                    (e.g., [HSW2005] 13)}
\tablenotetext{(b)}{Best fitting template spectrum type; see \S3.2 for descriptions.}
\tablenotetext{(c)}{Rest-frame Johnson K-band luminosity
                    in units of the sun's K-band luminosity, L$_{\odot, \rm K}$ = 8.7 $\times$ 10$^{24}$ W  
                    (Bessel 1979).}
\tablenotetext{(d)}{Monochromatic luminosity at rest-frame 5 $\mu$m in Watts, defined as $\nu$L$_{\nu}$(5 $\mu$m).}
\tablenotetext{(e)}{ Stellar mass derived by multiplying L$_{\rm K}$ by M$_{*}$/L$_{\rm K}$ =
                0.3, which is appropriate for a starburst (Bell \& de Jong 2001). }
\tablenotetext{(f)}{Estimated 8-1000 $\mu$m luminosity from Houck05.}
\end{deluxetable}

\clearpage
\begin{deluxetable}{ccccccccc}
\tablenum{5}
\tablecaption{Sub-millimeter Galaxies In The Lockman Hole East Region}
\tablewidth{0pc}
\tabletypesize{\footnotesize}
\tablehead{
\colhead{Source\tablenotemark{(a)} } & 
  \colhead{$z$\tablenotemark{(b)}} & 
  \colhead{Type\tablenotemark{(c)} } & 
  \colhead{L$_{\rm K}$\tablenotemark{(d)} } & 
  \colhead{L$_{\rm 5 \mu m}$\tablenotemark{(e)}} &
  \colhead{M$_{*}$\tablenotemark{(f)}  } &
  \colhead{L$_{\rm 20cm}$} & 
  \colhead{{\em q}\tablenotemark{(g)} } \\
  \colhead{[LE850]~$\#$} & \colhead{} & \colhead{} &
  \colhead{( 10$^{11}$~L$_{\odot, \rm K}$ )} & \colhead{( 10$^{37}$ W )} &
  \colhead{( 10$^{11}$~M$_{\odot}$ )} &
  \colhead{ ( 10$^{24}$ W~Hz$^{-1}$ )} & \colhead{}
}
\startdata
1     & 2.6 & cold & ~4.0 $\pm$ ~1.0  & ~1.4 $\pm$ 0.3 & 1.2 $\pm$ 0.3 &  3.1 $\pm$ 0.7 &  0.42 $\pm$ 0.10 \\
4     & 2.6 & cold & ~3.8 $\pm$ ~1.0  & ~1.3 $\pm$ 0.3 & 1.1 $\pm$ 0.3 &  1.3 $\pm$ 0.3 &  0.68 $\pm$ 0.51 \\ 
7     & 1.8 & cold & 12.2 $\pm$ ~4.7  & ~4.2 $\pm$ 1.5 & 3.7 $\pm$ 1.4 &  2.0 $\pm$ 0.8 &  0.36 $\pm$ 0.07 \\
8a    & 1.0 & cold & ~5.0 $\pm$ ~3.5  & ~1.8 $\pm$ 1.0 & 1.5 $\pm$ 1.1 &  1.2 $\pm$ 0.7 &  1.73 $\pm$ 0.24 \\
8b    & 3.0 & warm & 53.2 $\pm$ 12.5  & 18.4 $\pm$ 4.3 & $<$16.        &  4.1 $\pm$ 1.0 &  0.69 $\pm$ 0.14 \\
8c    & 0.9 & cold & ~1.4 $\pm$ ~0.6  & ~0.6 $\pm$ 0.3 & 0.4 $\pm$ 0.2 &$<$0.3          &  $>$0.81         \\
14a   & 2.4 & cold & ~6.7 $\pm$ ~2.0  & ~2.3 $\pm$ 0.7 & 2.0 $\pm$ 0.6 &  0.7 $\pm$ 0.3 &  0.36 $\pm$ 0.24 \\
14b   & 2.5 & cold & ~6.9 $\pm$ ~1.8  & ~2.4 $\pm$ 0.7 & 2.1 $\pm$ 0.5 &  1.6 $\pm$ 0.6 &  0.36 $\pm$ 0.15 \\
18    & 2.7 & warm & 11.4 $\pm$ ~3.0  & ~3.9 $\pm$ 1.0 & $<$3.4        &  1.1 $\pm$ 0.3 &  0.42 $\pm$ 0.16 \\
35    & 3.0 & cold & 20.3 $\pm$ ~4.8  & ~7.0 $\pm$ 1.7 & 6.1 $\pm$ 1.4 &  2.8 $\pm$ 0.7 &  0.44 $\pm$ 0.13 \\
\enddata
\tablenotetext{(a)}{Source number in Table 1 of Egami et al. (2004), where radio, mid-IR,
                    and sub-millimeter flux densities are given. Coordinates can be found via
                    SIMBAD using the designation LE850 followed by the source number (i.e.,
                    [LE850] 4).}
\tablenotetext{(b)}{Sources 8a, 14a, and 18 have spectroscopic redshifts. The other redshifts are photometric.
                    See footnotes {\em a} and {\em b} in Table 1 of Egami et al. (2004) for references.}
\tablenotetext{(c)}{Sources classified as ``warm'' possess power-law SEDs
                    in the {\em Spitzer} data and are likely AGN powered, whereas 
                    ``cold'' sources appear to be powered largely by star formation 
                    (e.g., bottom-right panel of Figure 7).}
\tablenotetext{(d)}{Rest-frame Johnson K-band luminosity in units of the sun's K-band luminosity, 
                    L$_{\odot, \rm K}$ = 8.7 $\times$ 10$^{24}$ W  (Bessel 1979).}
\tablenotetext{(e)}{Monochromatic luminosity at rest-frame 5 $\mu$m in Watts, defined as $\nu$L$_{\nu}$(5 $\mu$m).}
\tablenotetext{(f)}{Estimated stellar mass from multiplying L$_{\rm K}$ by M$_{*}$/L$_{\rm K}$
                    = 0.3, which is appropriate for a starburst (Bell \& de Jong 2001).}
\tablenotetext{(g)}{{\em q} $\equiv log(F_{\rm 24 \mu m}/F_{\rm 20cm})$.}

\end{deluxetable}


\clearpage
\begin{figure}
\figurenum{1}
\caption{Optically Invisible Radio Sources ({\it OIRSs}) from Higdon05 with
detections in the IRAC Shallow Survey (Eisenhardt et al. 2004). Optical
$B_W$ images are shown in the left-most column along with source index and
estimated photometric redshift (if known) as discussed in $\S$3.1. The middle-column
shows the same field with IRAC at 3.6 $\mu$m, along with 20cm radio contours starting
at 3 $\sigma$ from Higdon05. The third column shows the IRAC 3.6$\mu$m image without
radio contours. The fields are 20$''$ wide, which corresponds to
$\approx$175 kpc for $z = 1-3$. The source index refers to Table 2 of Higdon05 and 
Table 1 of this paper. A linear stretch is used for both optical and infrared 
grey-scale images. }
\end{figure}


\begin{figure}
\figurenum{2}
\caption{Optically Invisible Radio Sources ({\it OIRSs}) with no obvious 
counterparts in the IRAC Shallow Survey (Eisenhardt et al. 2004). Optical
$B_W$ images are shown in the left-most column while the corresponding
IRAC 3.6 $\mu$m images are shown to the immediate right. Radio 20cm contours
are shown in the middle IRAC 3.6 $\mu$m planes, starting at the 3 $\sigma$ level. The fields are
20$''$ in size. Also shown are the source index from Table 2 of Higdon05
and Table 1 of this paper. A linear stretch is used for both optical and infrared grey-scale images.}
\end{figure}


\begin{figure}
\figurenum{3}
\caption{Optically Invisible 24$\mu$m Sources ({\it OIMSs}) from Houck05. Optical $B_W$
images are shown in the left-most column along with redshifts derived from IRS
spectra. Opposite these are matching IRAC 3.6$\mu$m images from the IRAC Shallow Survey
(Eisenhardt et al. 2004). Each field is 20$''$ wide, 
which corresponds to $\approx$175 kpc for $z = 1-3$. The source index refers to Table 1
of Houck05 and Table 2 of this paper. A linear stretch is used for both
optical and infrared grey-scale images.}
\end{figure}

\clearpage
\begin{figure}
\figurenum{4}
\epsscale{1.0}
\plotone{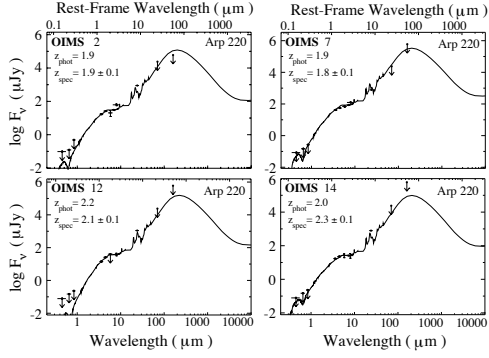}
\caption{Minimum-$\chi^{2}$ template fits for the {\it comb} and {\it sb}
OIMSs.  The Arp~220 SED template (solid line) gives photometric redshifts in good 
agreement with IRS derived $z_{\rm spec}$ ($\Delta z \approx 0.15$). 
Wavelengths are shown in the observer's frame on the bottom axis and 
in the rest-frame on the top axis.}
\end{figure}

\begin{figure}
\epsscale{1.}
\figurenum{5}
\plotone{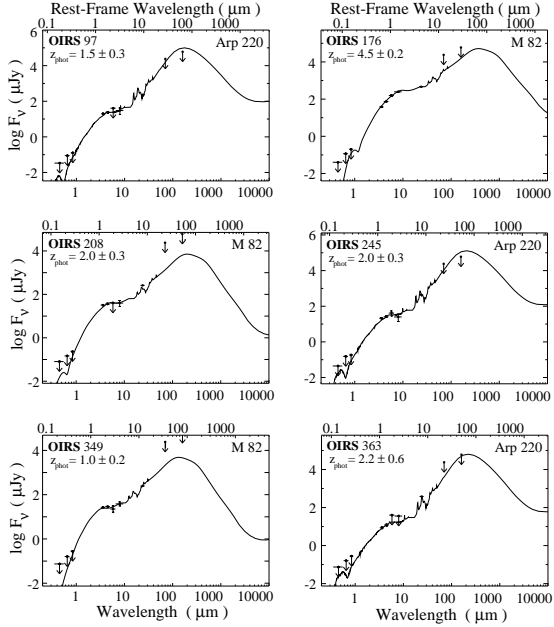}
\caption{Minimum-$\chi^{2}$ template fits for the six IRAC and MIPS 
detected OIRSs. The galaxy template
that best fits the data and the derived
$z_{\rm phot}$ are listed at the top of each panel.
Upper-limits are $3 \sigma$, and are depicted as arrows. 
Wavelengths are shown in the observer's frame on the bottom axis and 
in the rest-frame on the top axis.
}
\end{figure}

\clearpage
\begin{figure}
\epsscale{0.9}
\figurenum{6}
\plotone{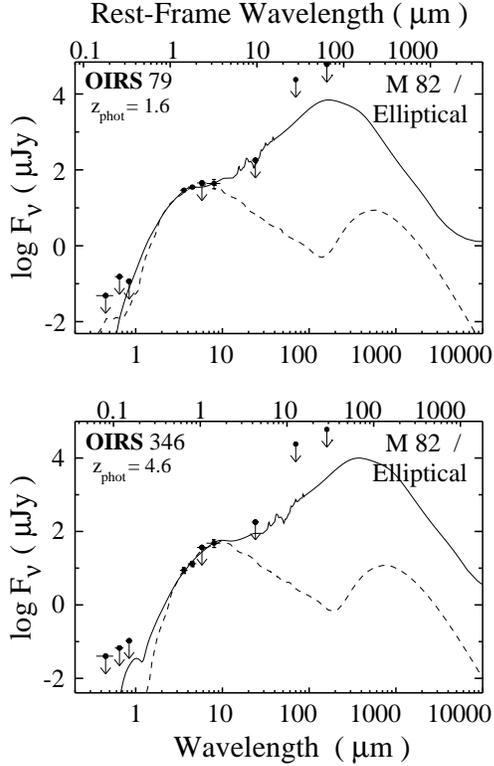}
\caption{Minimum-$\chi^{2}$ template fits for two OIRSs detected to $\ge 3 \sigma$
in at least three IRAC bands but undetected by MIPS. The axes
are defined as in Figures 4 and 5. The derived $z_{\rm phot}$ in each
instance is only an estimate. Similarly reasonable fits
using an elliptical template are depicted using
a dashed line. For either template choice, the estimated $z_{\rm phot}$ does
not change substantially.
Wavelengths are shown in the observer's frame on the bottom axis and 
in the rest-frame on the top axis.}
\end{figure}

\begin{figure}
\figurenum{7}
\epsscale{0.9}
\plotone{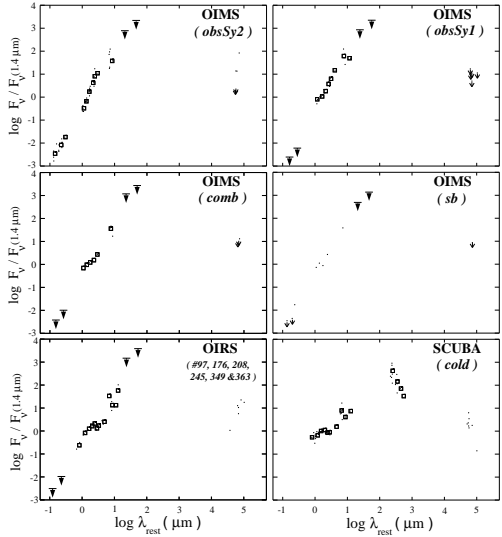}
\caption{Averaged rest-frame SEDs for the 
OIMSs grouped by the template galaxies that best fit their IRS spectra:
{\it obsSy2} (IRAS F00183-7111), {\it obsSy1} (Mrk~231), {\it comb} (Arp~220),
and {\it sb} (NGC~7714).  Also shown are averaged rest-frame SEDs for the six 
IRAC detected OIRSs with $z_{\rm phot}$ and the eight ``cold'' SMGs 
from Egami et al. (2004) and Ivison et al. (2004).  Each SED has been 
normalized to $\lambda_{\rm rest}$ = 1.4 $\mu$m. Individual data points (limits) are 
represented by dots (small arrows) while the unfilled squares represent binned 
averages. Large arrows represent averaged limits. Note that the {\it sb}-like 
SED in the middle-right panel consists of a single source (OIMS $\#$2) and has not
been rebinned. }
\end{figure}

\clearpage
\begin{figure}
\figurenum{8}
\epsscale{0.8}
\plotone{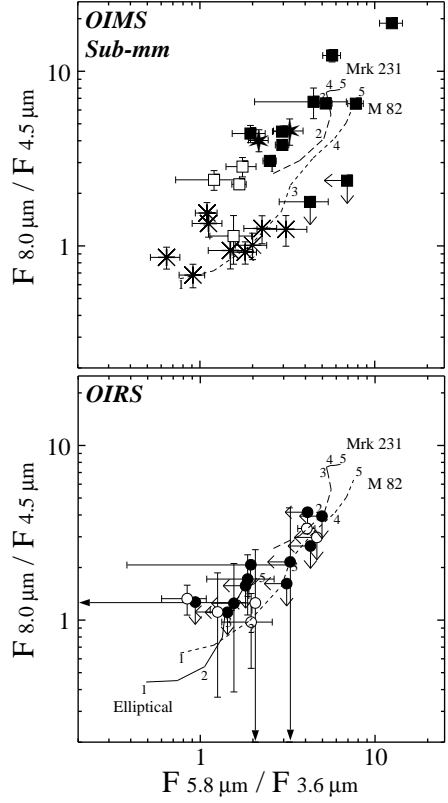}
\caption{IRAC two-color diagrams. 
The top panel shows OIMSs (filled squares if their IRS spectra are
{\it obsSy1} or {\it obsSy2}-like and empty squares if {\it sb} or {\it comb}-like)
along with ten SMGs in the Lockman Hole East region
(asterisks for ``cold''  and six-pointed stars for ``warm'' sources) from Egami et al. (2004). 
Tracks corresponding to Mrk~231 and a dusty M~82 template (extra A$_{\rm V}$ = 
2.5) for $z = 1-5$ are over-plotted. OIMS \#17 was not observed
at 4.5 $\mu$m and 8.0 $\mu$m and is not shown. The lower panel
shows the OIRSs from Table 1 as circles. Those with $z_{\rm phot}$ 
are unfilled. A track corresponding to a model elliptical
galaxy for $z = 1-5$ is also shown.
}
\end{figure}

\begin{figure}
\figurenum{9}
\epsscale{1.0}
\plotone{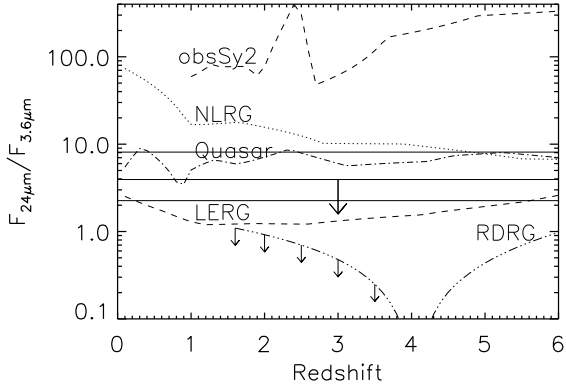}
\caption{ The rest-frame $F_{\rm 24 \mu m}/F_{\rm 3.6 \mu m}$ ratio
as a function of redshift for five sources known to contain powerful
AGN: (1) the averaged {\it obsSy2} SED shown in
Figure 7, (2) the NLRG 3C459, (3) the quasar 3C273, (4) the LERG
3C293, and (5) the ``red and dead'' radio galaxy LBDS 53W091 (Stern et al.
2006). The solid horizontal line at $F_{\rm 24 \mu m}/F_{\rm 3.6 \mu m} = 3.9$ indicates the  
$3 \sigma$ upper-limit for the twelve IRAC detected OIRSs without 24 $\mu$m detections,
defined by dividing the stack-averaged upper-limit at 24 $\mu$m by the average
3.6 $\mu$m flux density. The two horizontal lines above and below this show the range in  
$F_{\rm 24 \mu m}/F_{\rm 3.6 \mu m}$ for these OIRSs defined by dividing the 
24 $\mu$m upper-limit by their individual 3.6 $\mu$m flux densities. These
OIRSs appear to have SEDs most consistent with LERGs and ``red and dead''
radio galaxies like LBDS 53W091 over a wide redshift range.}
\end{figure}

\clearpage
\begin{figure}
\figurenum{10}
\epsscale{1.2}
\plotone{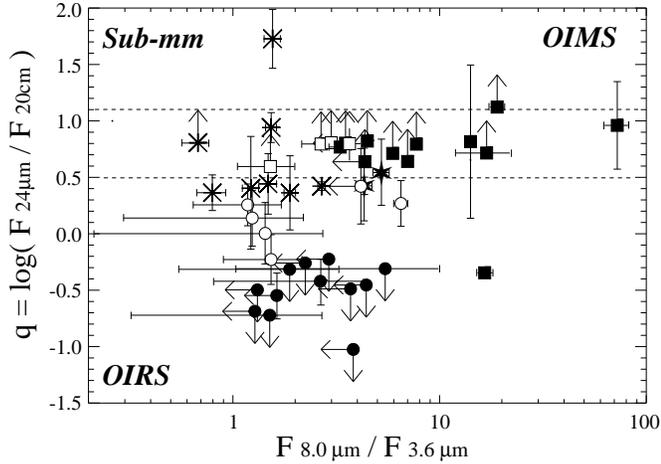}
\caption{A radio-infrared two-color diagram showing OIMSs, IRAC
detected OIRSs, and the ten SMGs in the Lockman
Hole east region observed by Egami et al. (2004). The AGN-dominated
OIMSs (i.e., {\it obsSy1} and {\it obsSy2}) are depicted as filled squares
while the OIMSs with a significant starburst component ({\it comb} and
{\it sb}) are depicted as empty squares. The OIRSs are shown as unfilled circles
if they have $z_{\rm phot}$ estimates, and filled circles if no redshift
estimates are available. Similarly, ``cold'' starburst dominated SMGs
are shown as asterisks, while the ``warm'' AGN dominated SMGs
are shown as six-pointed stars. The dashed horizontal 
lines show the limits of the radio-infrared correlation determined 
by Appleton et al. (2004) at 24 $\mu$m and 20cm ({\it q} = 0.8 $\pm$ 0.3). 
OIRS $\#$278 and OIMS $\#$17 could not be plotted.}
\end{figure}

\begin{figure}
\figurenum{11}
\epsscale{1.2}
\plotone{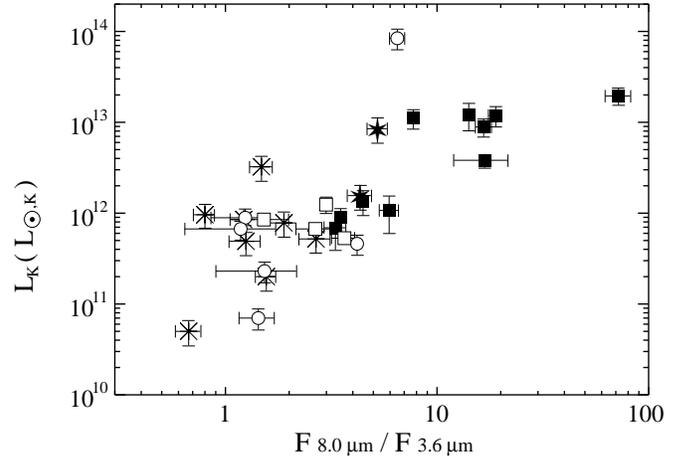}
\caption{Rest-frame K-band luminosity as a function of observer-frame 
$F_{\rm 8.0 \mu m}/F_{\rm 3.6 \mu m}$ flux density ratio for OIMSs
(filled squares for {\it obsSy1} and {\it obsSy2}-like systems, empty
squares for {\it sb} and {\it comb}-like systems) and ten SMGs
from the Lockman Hole East region observed by Egami et al. (2004).
Starburst-like ``cold'' sources are shown as asterisks while AGN-like
``warm'' sources are depicted as six-pointed stars. The six IRAC detected OIRSs with
photometric redshifts are also shown as empty circles.}
\end{figure}

\end{document}